\documentclass[preprint]{aastex61}
\usepackage{textcomp}
\usepackage{amsmath}
\received{xxx}
\revised{xxx}
\accepted{xxx}

\shorttitle{SN\,2017eaw molecule and dust}
\shortauthors{Tinyanont et al.}

\begin{document}

\title{Supernova 2017eaw: molecule and dust formation from infrared observations}

\correspondingauthor{Samaporn Tinyanont}
\email{st@astro.caltech.edu}

\author{Samaporn Tinyanont}
\affiliation{Division of Physics, Mathematics and Astronomy, California Institute of Technology, 1200 E. California Blvd., Pasadena, CA 91125, USA}
\nocollaboration

\author{Mansi M Kasliwal}
\affiliation{Division of Physics, Mathematics and Astronomy, California Institute of Technology, 1200 E. California Blvd., Pasadena, CA 91125, USA}

\author{Kelsie Krafton}
\affiliation{Department of Physics and Astronomy, Louisiana State University, Baton Rough, LA 70803, USA}

\author{Ryan Lau}
\affiliation{Institute of Space \& Astronautical Science, Japan Aerospace Exploration Agency, 3-1-1 Yoshinodai, Chuo-ku, Sagamihara, Kanagawa 252-5210, Japan}

\author{Jeonghee Rho}
\affiliation{SETI Institute, 189 N. Bernardo Ave, Suite 200, Mountain View, CA 94043, USA}
\affiliation{SOFIA Science Center, NASA Ames Research Center, MS 232, Moffett Field, CA 94035, USA}

\author{Douglas C Leonard}
\affiliation{Department of Astronomy, San Diego State University, San Diego, CA 92182, USA}

\author{Kishalay De}
\affiliation{Division of Physics, Mathematics and Astronomy, California Institute of Technology, 1200 E. California Blvd., Pasadena, CA 91125, USA}

\author{Jacob Jencson}
\affiliation{Division of Physics, Mathematics and Astronomy, California Institute of Technology, 1200 E. California Blvd., Pasadena, CA 91125, USA}

\author{Dimitri Mawet}
\affiliation{Division of Physics, Mathematics and Astronomy, California Institute of Technology, 1200 E. California Blvd., Pasadena, CA 91125, USA}
\affiliation{Jet Propulsion Laboratory, California Institute of Technology, Pasadena, CA 91109, USA}

\author{Maxwell Millar-Blanchaer}
\affiliation{Jet Propulsion Laboratory, California Institute of Technology, Pasadena, CA 91109, USA}
\affiliation{Hubble Fellow}

\author{Ricky Nilsson}
\affiliation{Division of Physics, Mathematics and Astronomy, California Institute of Technology, 1200 E. California Blvd., Pasadena, CA 91125, USA}
\affiliation{Jet Propulsion Laboratory, California Institute of Technology, Pasadena, CA 91109, USA}

\author{Lin Yan}
\affiliation{Division of Physics, Mathematics and Astronomy, California Institute of Technology, 1200 E. California Blvd., Pasadena, CA 91125, USA}

\author{Robert D Gehrz}
\affiliation{Minnesota Institute for Astrophysics, School of Physics and Astronomy, University of Minnesota, 116 Church Street, S. E., Minneapolis, MN 55455, USA}

\author{George Helou}
\affiliation{Caltech/IPAC, Mailcode 100-22, Pasadena, CA 91125, USA}

\author{Schuyler D Van Dyk}
\affiliation{Caltech/IPAC, Mailcode 100-22, Pasadena, CA 91125, USA}

\author{Eugene Serabyn}
\affiliation{Jet Propulsion Laboratory, California Institute of Technology, 4800 Oak Grove Dr, Pasadena, CA 91109, USA}

\author{Ori D Fox}
\affiliation{Space Telescope Science Institute, 3700 San Martin Dr, Baltimore, MD 21218, USA}

\author{Geoffrey Clayton}
\affiliation{Department of Physics and Astronomy, Louisiana State University, Baton Rough, LA 70803, USA}




\begin{abstract}
We present infrared (IR) photometry and spectroscopy of the Type II-P SN\,2017eaw and its progenitor in the nearby galaxy NGC\,6946. 
Progenitor observations in the Ks band in 4 epochs from 1 year to 1 day before the explosion reveal no significant variability in the progenitor star greater than 6\% that last longer than 200 days.
SN\,2017eaw is a typical SN II-P with near-IR and mid-IR photometric evolution similar to those of SNe\,2002hh and 2004et, other normal SNe II-P in the same galaxy.
Spectroscopic monitoring between 389 and 480 days post explosion reveals strong CO first overtone emission at 389 d, with a line profile matching that of SN\,1987A from the same epoch, indicating $\sim 10^{-3} \, M_{\odot}$ of CO at 1,800 K. 
From the 389 d epoch until the most recent observation at 566 d, the first overtone feature fades while the 4.5 $\mu$m excess, likely from the CO fundamental band, remains.
This behavior indicates that the CO has not been destroyed, but that the gas has cooled enough that the levels responsible for first overtone emissions are no longer populated. 
Finally, the evolution of \textit{Spitzer} 3.6 $\mu$m photometry shows evidence for dust formation in SN\,2017eaw, with a dust mass of $10^{-6}$ or $10^{-4}\,M_{\odot}$ assuming carbonaceous or silicate grains respectively. 
\end{abstract}




\section{Introduction}
Massive stars ($M \gtrsim 8 \,M_{\odot}$) conclude their evolution in a core-collapse supernova (CCSN) when the nuclear fuel in their core is exhausted.
The most common CCSNe are of Type II-P, whose defining features are a plateau in their light curves where the bolometric luminosity remains constant for $\sim$100~days and strong hydrogen features in all phases of their spectral evolution (see e.g. \citealp{filippenko1997} for SN types).
The persisting presence of hydrogen points to progenitor stars with hydrogen remaining in the stellar envelope at the time of core-collapse.
Archival pre-explosion observations of a number of nearby SNe II-P have indeed revealed hydrogen-rich red supergiant (RSG) progenitor stars (\citealp{smartt2009} and references therein).
While SNe II-P are the most common---and most well understood---among subtypes of CCSNe, there are several aspects of the explosion that remain to be understood, both theoretically and observationally.
Examples of unanswered questions regarding SNe II-P include the landscape of the circumstellar medium (CSM) around their RSG progenitors and the chemical evolution and dust formation in their ejecta. 
Some SNe II-P show signatures of shock interaction with a circumstellar medium (CSM), even though the typical RSG progenitors are expected to have steady wind driven mass loss with no dense CSM at the time of core-collapse. 
These signs of shock interaction include weak X-ray and radio emission from the shocked material and infrared (IR) emission from dust, that is either pre-existing or newly formed. 
Another aspect of SNe II-P that is not yet well understood is the chemical evolution of their ejecta. 
CCSNe are a promising source of dust in the early universe, making it crucial to understand how molecules and dust grains evolve in their ejecta.  
While chemical evolution models of the ejecta of SNe II-P have existed for decades, observational data (mostly IR) required to test them remain sparse due to the limited number of nearby normal SNe II-P that can be observed in great details out to late times. 

While we expect SNe II-P in a pristine environment in comparison to SNe with stripped-envelope progenitors, recent observations, especially in the infrared (IR), show several SNe II-P with weaker signs of interactions with a CSM likely ejected within decades to centuries before the SN.
In the strongest case of interaction with a dense CSM, one can observe narrow recombination lines from the CSM gas photoionized by high energy photons from the interaction region \citep[spectral Type IIn;][]{schlegel1990}.
For SNe II-P, however, the narrow lines are not present and the main signatures of the interaction are X-ray and radio emissions coming from the shocked CSM gas, with lower luminosity indicating a less dense CSM in comparison to that of SNe IIn (e.g. SNe\,2013ej, \citealp{Chakraborti2013}; 2011ja, \citealp{Chakraborti2016}).   
CSM interactions also leave spectroscopic imprints, such as a high velocity absorption feature of H$\alpha$ and He I 1.083 $\mu$m (\citealp{chugai2007}) and asymmetric and multi-peaked hydrogen lines associated with a toroidal CSM (e.g. SNe\,2007od, \citealp{andrews2010}; 2011ja, \citealp{andrews2016}). 
Finally, the CSM interaction can trigger new dust formation and heat pre-existing dust, both of which emit in the thermal IR.

The CSM interaction can trigger dust formation by generating a reverse shock propagating back into the ejecta, creating a cold dense shell (CDS) between the forward and reverse shocks. 
Conditions in the CDS are suitable for dust formation.
In some SNe, CDS dust formation can happen early ($\lesssim 300$ d), before the outer ejecta have cooled enough for dust formation (although recently \citealp{sarangi2018b} have argued that before 380 d, high energy photons from the shock interaction region inhibit dust formation in the CDS).
Early dust formation, likely in the CDS, has been observed in many interacting SNe, e.g. SNe\,2005ip, \citep[IIn][]{smith2009b,fox2009}; 2006od, \citep[II-P][]{andrews2010}; 2010jl, \citep[IIn][]{gall2014}; 2011ja \citep[II-P][]{andrews2016, tinyanont2016}. 
Nearby SNe 2004dj and 2004et both showed re-brightening in their IR light curves attributable to dust formation \citep{szalai2011, kotak2009,meikle2011,fabbri2011}.
In SN\,2004et, the dust formation at $\sim$1,000 days post-explosion likely occurred in the CDS between the SN forward shock and the reverse shock generated by a delayed CSM interaction. 
\cite{kotak2009} showed that the SED from 200 to 500 d post explosion exhibited strong signatures of carbon monoxide (CO) and silicon monoxide (SiO) at 4.67 and 8 $\mu$m along with a broad silicate grain feature around 9.7 $\mu$m, suggesting a mixture of carbonaceous and silicate grains formation.
Most of these observations were enabled by \textit{Spitzer} Space Telescope \citep{werner2004,gehrz2007}/Infrared Array Camera \citep[IRAC;][]{fazio2004} observations and, for a limited number of CCSNe, by the cold mission instruments, Infrared Spectrograph \citep[IRS;][]{houck2004} and the Multiband Imaging Photometer for \textit{Spitzer} \citep[MIPS;][]{rieke2004}. 
These observations are sensitive to the thermal emission from dust with temperature of 200-1,000 K.
We note here that the CSM dust can create an IR light echo from the peak SN light without any CSM-shock interaction.
However, observations have shown that IR light echoes cannot explain the large amount of IR radiation seen at late times, even in SN\,2011dh where the IR light curve faded quickly in comparison to other CCSNe \citep{helou2013}.


In addition to observations of a handful of SNe during its cold mission, \textit{Spitzer} observed multiple SNe during the warm mission when only the IRAC 3.6 and 4.5 $\mu$m imaging channels were available.
While the full spectral coverage, especially around the 9.7 $\mu$m silicate feature, is not available, warm \textit{Spitzer} studies have shown a diversity in the IR light curve evolution among SNe II-P.
Our previous work \citep{tinyanont2016} presented a \textit{Spitzer} survey of 36 CCSNe in nearby galaxies out to 20 Mpc showing that IR emission later than 100~d is common among CCSNe, with some SNe II-P showing signs of CSM interactions (see \citealp{szalai2018} for a more recent compilation of \textit{Spitzer} observations of SNe). 
The most unique among our sample were SN\,2011ja with bright and almost constant IR luminosity out to $\sim$1000~d and SN\,2013ej with IR rebrightening.
Both SNe showed signs of CSM interactions from X-Ray observations \citep{Chakraborti2013, Chakraborti2016}, optical spectroscopy \citep{andrews2016, Mauerhan2017}, and, for SN\,2013ej, spectropolarimetry \citep{Mauerhan2017}.
SN\,2011ja likely formed dust very early at 105 d post explosion \citep{andrews2016}. 
These observations demonstrate the range of epochs at which the CSM interactions and/or dust formation begin around SNe II-P. 

Dust grains can form not only in the CDS created by the CSM interaction, but also in the ejecta of the SN itself after they have sufficient time to cool. 
This is arguably the more important channel of dust formation because it can operate whether or not the SN has a CSM interaction.
To understand dust formation in the ejecta, it is crucial to have a realistic chemical evolution model. 
Several models in the past have relied on simplistic Classical Nucleation Theory (CNT) or Kinetic Nucleation Theory (KNT) in which dust formation is parameterized in some way to simplify the calculation (see \citealp{sluder2018, sarangi2018} for a summary of different classes of chemical evolution models). 
In the past decade, modelers have started to employ Molecular Nucleation Theory (MNT), which simulates molecule and dust formation in the SN ejecta only using a realistic network of chemical reactions. 
The MNT models are the only ones that explicitly simulate molecule and dust evolution simultaneously. 
Molecules are crucial for dust formation because they are effective at cooling the ejecta to temperatures suitable for dust condensation and can act as seed nuclei.
Some molecules are direct precursor species to dust grains.
For example, SiO is a building block for silicate grains. 
The summary of observational data on CO and SiO mass evolution in a few SNe in comparison to a 15 $M_\odot$ model can be found in Figures 3 and 4 in \cite{sarangi2013} respectively.
The same comparison for observed dust mass in 12 SNe and 4 different progenitor models can be found in Figure 10 from \cite{sarangi2015}.
The observations required to measure molecule and dust mass are difficult, and are sparse as a result.
For molecule formation in SN ejecta, only $\sim$10 observations have been reported (see the summaries by \citealp{gerardy2002} and \citealp{sarangi2018}, and references therein).
Additional observations are required to bridge the gap between the small dust masses of $10^{-4} - 10^{-3} \, M_\odot$ inferred at few hundred days post explosion from near to mid-IR observations \citep[e.g][]{szalai2013, tinyanont2016} and the larger dust masses of 0.01-0.1 $M_\odot$ inferred in supernova remnants Cas A \citep{rho2008, barlow2010}, Crab Nebula \citep{gomez2012}, Sgr A East \citep{lau2015}, and in SN\,1987A \citep{indebetouw2014, matsuura2015} inferred from far-IR and sub-mm observations.



SN\,2017eaw is the most recent nearby event for which detailed late-time IR observations, required to investigate the issues described above, were possible.
SN\,2017eaw was discovered on 2017 May 14 in NGC\,6946 as its tenth SN in the past century \citep{wiggins2017}.
Early spectroscopic observations showed that it was a typical young SN II-P with a spectrum similar to that of SN\,1999gi at 3.8 days post-explosion \citep{tomasella2017}.
Early radio observations immediately post-detection resulted in non-detections at 1.39, 5.1, and 15 GHz by the Giant Metrewave Radio Telescope (GMRT), the electronic Multi-Element Remotely-Linked Interferometer Network (e-MERLIN), and the Arcminute Microkelvin Imager Large Array (AMI-LA) respectively. 
At 1.39 GHz, \cite{nayana2017a} reported an early non-detection on 2017 May 18 with an upper limit of 114 $\mu$Jy. 
Thirty six days later on 2017 June 23, the SN was detected at 230$\pm$68 $\mu$Jy \citep{nayana2017b}. 
At 5.1 GHz, \cite{argo2017a} reported an early non detection on 2018 May 20 with an upper limit of 65 $\mu$Jy/beam.
Subsequent observations on May 29.55, 30.81, and June 1.71 detected a rising flux of 280$\pm$25, 350$\pm$51, 427$\pm$23 $\rm\mu$Jy respectively \citep{argo2017b}.
Lastly, AMI-LA observations at 15 GHz on May 15, 17, and 20 did not yield any detections with upper limits of 0.5 - 1 mJy \citep{bright2017,mooley2017}.
No AMI-LA observations have been reported during the epochs in which the SN was detected at 1.39 and 5 GHz. 
This increasing radio flux may be a sign of the SN shock running into the denser part of the CSM.
X-ray observations by Swift on May 14 \citep{kong2017} and NuStar on May 21 \citep{grefensetette2017} revealed a rising X-ray flux, also indicative of CSM interactions. 
The  Nuclear Spectroscopic Telescope Array (NuStar; \citealp{harrison2013}) observations showed a hard spectrum with photons detected up to 30 keV and also revealed a shock heated ionized Fe line at 6.65 keV. 
SN\,2017eaw has also been observed with \textit{Spitzer}/IRAC in the 3.6 and 4.5 $\mu$m bands as part of the ongoing SPitzer InfraRed Intensive Transients Survey (SPIRITS; \citealp{Kasliwal2017}).
In addition to these potential signs of CSM interactions, optical and near-IR observations of this SN are consistent with those of other SNe with early dust formation.
\cite{tsvetkov2018} published optical photometry of the SN until 200 d post-explosion along with a preliminary comparison to different light curve models. 
They noted the photometric similarity between this SN and SN\,2004et, an aforementioned SN II-P in the same galaxy. 
\cite{rho2018} reported Gemini near-IR spectroscopic follow-up from 22 to 205 d post discovery showing an emerging CO first overtone emission at $\sim 2.3 \, \mu$m, a clear signature of CO formation.
They also reported a rising red continuum in the K band that could be coming from hot dust. 

Here we present and discuss results from IR observations of SN\,2017eaw and its progenitor.
\textsection \ref{sec:observation} summarizes archival near-IR imaging of the progenitor star and IR photometry and spectroscopy of the SN with both ground-based telescopes and \textit{Spitzer}.
In \textsection \ref{sec:prog_phot}, we report near-IR photometry of the progenitor star of SN\,2017eaw detected in ground-based imaging, including one epoch by Keck/MOSFIRE on 2017 May 9, 5 days before the discovery of the SN. 
We present IR light curves and spectroscopy of the SN in \textsection \ref{sec:sn_phot}, \ref{sec:sn_spec}. 
We analyze the data in \textsection \ref{sec:analysis} and discuss the progenitor non-variability in the K band (\textsection \ref{sec:prog_sed}), the IR photometry in comparison to other SNe II-P (\textsection \ref{sec:LC_spec_comparison}), and SED (\textsection \ref{sec:sed}) and CO line profile and its evolution (\textsection\ref{sec:co_evo}, \ref{sec:co_profile}).
A summary and conclusions are in \textsection \ref{sec:discussion}.

\section{Observations}\label{sec:observation}
\subsection{Progenitor infrared photometry}\label{sec:prog_phot}
\begin{figure}
	\includegraphics[width=0.71\textwidth]{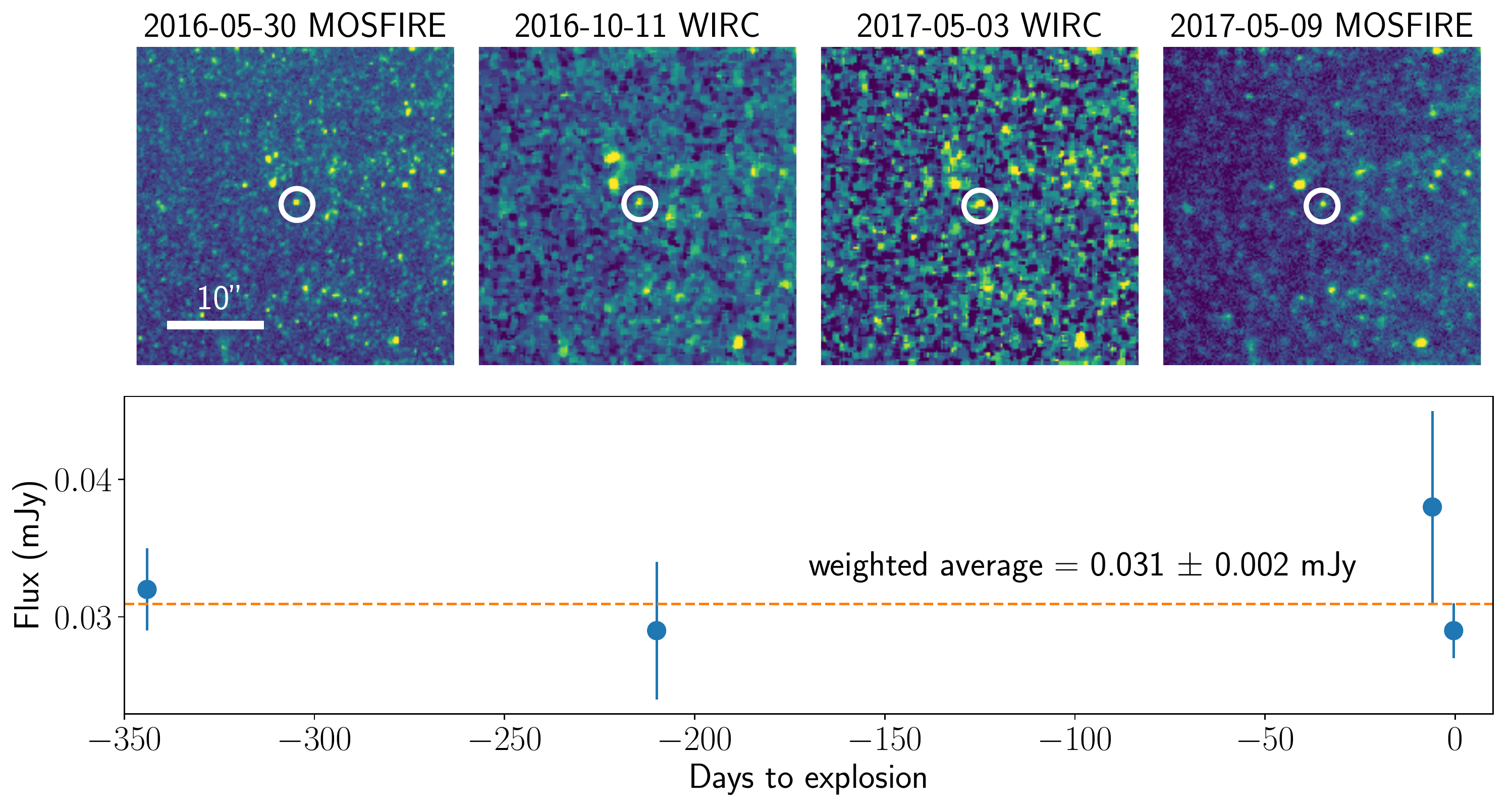} \hfill
    \includegraphics[width = 0.28\textwidth]{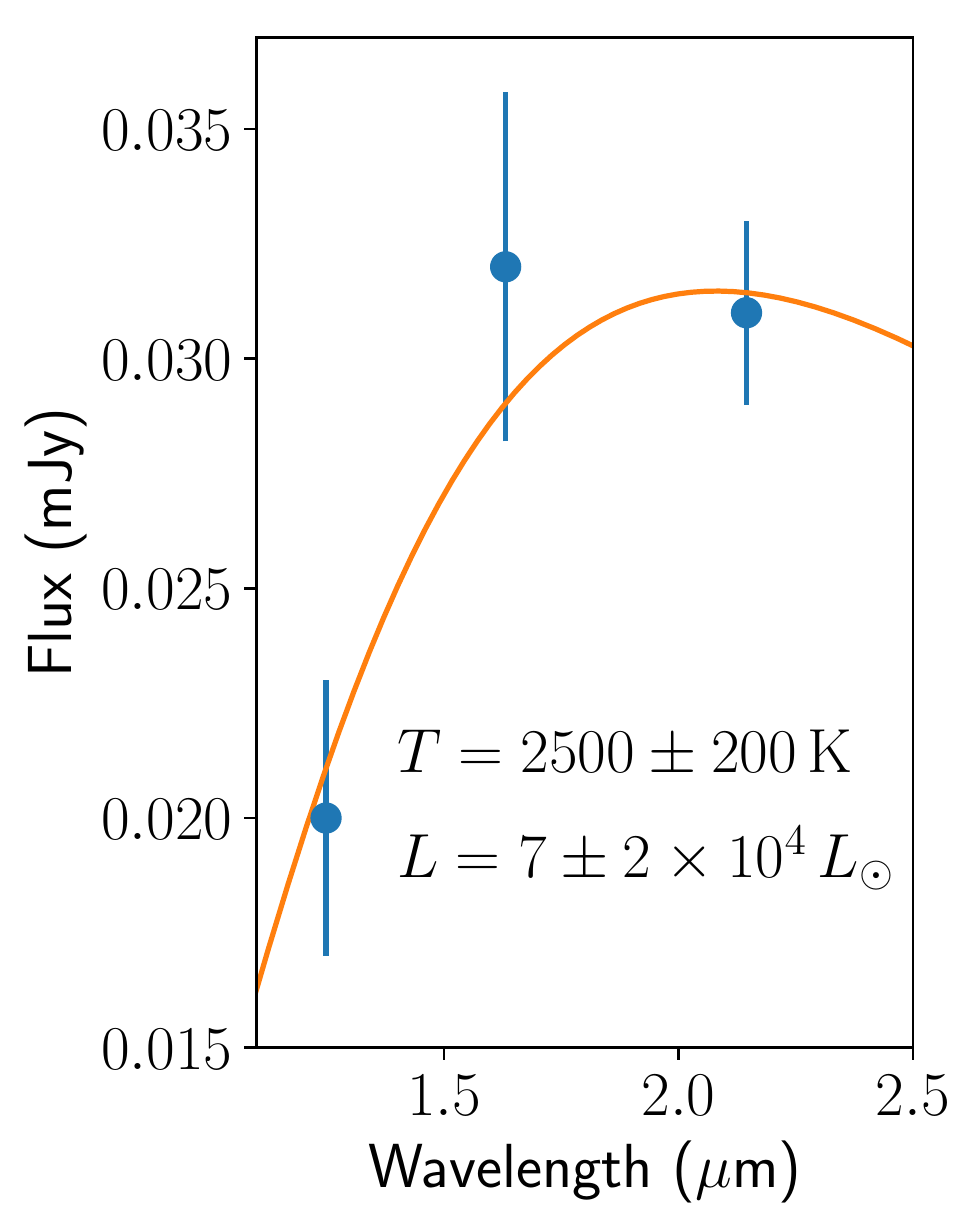}
    \caption{\textit{Left} Images of the progenitor of SN\,2017eaw from a year to $\sim$1 day before explosion (5 days before first detection) taken in the Ks band with Keck/MOSFIRE and P200/WIRC. The progenitor is encircled in each image. WIRC images have been smoothed using a median filter for visualization. Photometry from these images (shown below the images) revealed no significant variability of the progenitor at a 6\% level.
    \textit{Right} The near-IR SED of the progenitor star corrected for foreground extinction \citep{schlafly2011}. Data in the J and H bands are from 2016 October 11 while the Ks band is the weighted average over all 4 epochs shown on the left. Assuming $d = 7.72$ Mpc, the best blackbody parameters fitted to the SED are $T = 2{,}500 \pm 200 \, \rm K$ and $L = 7\pm2 \times 10^4 \, L_{\odot}$. These parameters are consistent with a RSG progenitor.}
    \label{fig:progenitor_image}
\end{figure}

NGC\,6946 is a nearby galaxy \citep[$d=7.72$ Mpc;][]{anand2018}, extremely prolific at producing SNe.
Numerous pre-explosion observations of SN\,2017eaw's site, both from the ground and space, are available partly as a result of efforts to follow the temporal evolution of SNe\,2002hh, 2004et, and 2008S.
The progenitor star has been detected in archival \textit{Hubble} and \textit{Spitzer} images, with a spectral energy distribution (SED) consistent with that of a dusty red supergiant star \citep{kilpatrick2018, vandyk2017,khan2017}.
As part of the follow-up campaign for SPIRITS, we have imaged the SN site with the Wide field InfraRed Camera (WIRC; \citealp{wilson2003}) on the 200-inch telescope at Palomar Observatory (P200 hereafter) in the J, H, and Ks bands on 2016 October 11, 7 months before the explosion, and in the Ks band on 2017 May 3, 11 days before the SN was first detected.
We also imaged the SN site in the Ks band using the Multi-Object Spectrometer For Infra-Red Exploration (MOSFIRE; \citealp{mclean2012}) at Keck Observatory on 2016 May 30 and 2017 May 9.
The second epoch was only 5 days before the SN was first detected.  
In both cases, data were taken with dithering patterns that send the galaxy in and out of the field of view to measure and subtract the sky background. 
Science images were dark subtracted and flattened using a flat field image obtained from median combining dithered sky images with sources masked out.
The galaxy light at the SN location is low in these bands, and we did not attempt any galaxy light subtraction. 
A point source is detected at the location of the SN in all observations, as shown in Fig.~\ref{fig:progenitor_image} \textit{top left}. 
Aperture photometry with sky annulus subtraction was obtained with the zero point determined using $\sim$100 stars in the field of view with magnitudes from Two Micron All-Sky Survey (2MASS; \citealp{milligan1996}, \citealp{skrutskie2006}). 
The results are listed in Table \ref{table:NIR_phot} and plotted in Fig.~\ref{fig:progenitor_image} \textit{bottom left}. 
We did not detect any significant variability, with a timescale greater than 200 days, of the progenitor star in the Ks band. 

\subsection{Explosion epoch}\label{sec:sn_boom}
\cite{tsvetkov2018} published optical photometry in UBVRI bands with data spanning from discovery (2018 May 14) until 208 days post-discovery. 
They fitted a suite of light curve models to the data and derived an explosion date of 2017 May 4, 10 days before the first detection.
We note that this is inconsistent with our progenitor observation on 2017 May 9, which showed no increase in flux from the progenitor at that epoch. 
We instead fitted a low order polynomial to their R and I bands light curves, which best capture the rise.
We constrain the explosion date to 2017 May 10, consistent with our progenitor observation and also with the spectroscopic age constraint (+3.8 d on May 14; \citealp{tomasella2017}).
This simple polynomial fitting neglects the shock break out phase, which typically only lasts $\sim$1 day. 
We take 2017 May 10 as the explosion date of SN\,2017eaw throughout this paper. 

\subsection{Supernova photometry}\label{sec:sn_phot}
We obtained photometry of SN\,2017eaw in the near-IR J, H, and Ks bands using P200/WIRC and Keck/MOSFIRE in 9 epochs, spanning 31 to 490 days post explosion, with 7 epochs during the nebular phase. 
The process for data reduction and aperture photometry was the same as in the last section. 
All near-IR photometry are presented in Table \ref{table:NIR_phot}, including those reported in \cite{arkharov2017} and \cite{rho2018}.
The 2MASS magnitudes were converted to flux densities using zero magnitude flux densities of 1594, 1024, and 666.7\,Jy for J, H, and Ks bands respectively \citep{cohen2003}. 

\begin{figure}
\centering
	\includegraphics[width=0.8\textwidth]{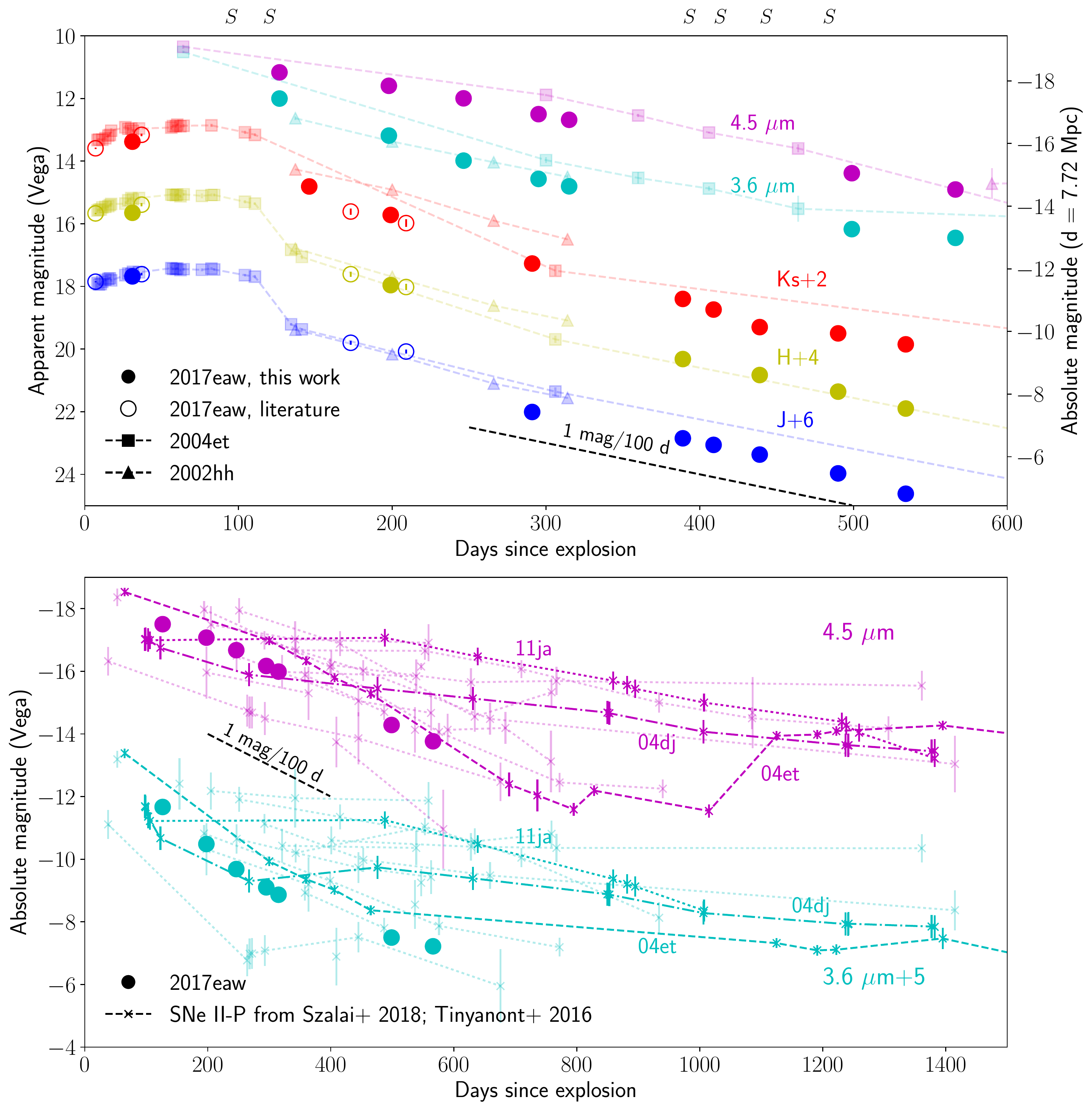}
   	\caption{\textit{Top:} Near-IR photometry of SN\,2017eaw in the J, H, Ks bands and the IRAC 3.6 and 4.5 $\mu$m bands compared with photometry of SN\,2004et and SN\,2002hh. Filled symbols for SN\,2017eaw are our data while open symbols are from \cite{arkharov2017} and \cite{rho2018}. The JHKs photometry for SN\,2004et are from \cite{maguire2010}. \textit{Spitzer} photometry are from \cite{fabbri2011}. 
    SN\,2002hh data are from \cite{pozzo2006}. Note that the 3.6 $\mu$m data for this SN is from L' band ground based observations. The ``S'' marks above the axis indicate epochs for which we obtained spectroscopy.
    \textit{Bottom:} SN\,2017eaw \textit{Spitzer} photometry compared to all other SNe II-P photometry as aggregated by \cite{szalai2018}. No offsets between SNe are applied here, but all 3.6 $\mu$m magnitudes are shifted by 5 mags for visualization. 
    SNe 2004dj, 2004et, and 2011ja are highlighted for comparison.
    All magnitudes here are in Vega system. 
    The 1 mag/100 d decline rate expected from light curves powered by radioactive decay of $\rm ^{56}Co$ is plotted in both subplots. }
    \label{fig:compare_17eaw}
\end{figure}

SN\,2017eaw was observed with the \textit{Spitzer} InfraRed Array Camera (IRAC; \citealp{fazio2004}) at 3.6 $\mu$m and 4.5 $\mu$m in 6 epochs: 126, 197, 246, 295, 315, 498, and 566 days post-explosion (PID\,13053, PI Kasliwal; PID\,13239, PI Krafton).
We used a stack of archival pre-explosion \textit{Spitzer} images to estimate and remove the galaxy background and nearby source contamination.\footnote{Archival \textit{Spitzer} images used for background subtraction came from the following PIDs and PIs: 60071, 70008, 80131, 90178 PI Andrews; 80015, 10081, 11084 PI Kochanek;  10136, 11063, 80196, 13053 PI Kasliwal; 10002 PI Sugerman.}
Archival images were rotated and aligned based on the sky coordinates supplied in \textit{Spitzer} data, then median combined.
In IRAC images in both channels of the 126 d epoch and the 3.6 $\mu$m channel of the 215 d epoch, columns of low counts due to a saturating star in the field of view crossed the SN.
To remove these low count columns, we fitted a Gaussian profile across each column and added the missing flux back in. 
We conducted aperture photometry on the background subtracted images and applied appropriate aperture corrections as given by the IRAC instrument handbook. 
\textit{Spitzer} photometry are listed in Table \ref{table:NIR_phot}.
\textit{Spitzer} fluxes were converted to magnitudes for plotting purposes (Fig.~\ref{fig:compare_17eaw}) using the zero magnitude fluxes of 280.9 and 179.7 Jy for the 3.6 and 4.5 $\mu$m channels respectively.
Fig. \ref{fig:compare_17eaw} shows the near-IR light curves of SN\,2017eaw in comparison with those of other well studied SNe II-P. 
The \textit{top} panel shows light curves of the J, H, Ks, 3.6, and 4.5 $\mu$m bands of SN\,2017eaw in comparison to those of SNe II-P 2002hh and 2004et in the same galaxy. 
The \textit{bottom} panel shows only \textit{Spitzer} data points compared with all SNe II-P observed with \textit{Spitzer} \citep[photometry taken from][]{szalai2018}.
Throughout the paper, we assume the distanc to NGC\,6946 of 7.72 Mpc based on the Tip of the Red Giant Branch (TRGB) technique \citep{anand2018} (also used by \citealp{rho2018}).

\begin{table}
\centering
\scriptsize
\caption{Near-infrared photometry}
\label{table:NIR_phot}
\begin{tabular}{cccccccccccccc} \toprule
Date & MJD& Epoch & $F_{J}$ & $\sigma_{F_{J}}$ & $F_{H}$ & $\sigma_{F_{H}}$ & $F_{K_s}$ & $\sigma_{F_{K_s}}$ & $F_{[3.6]}$ & $\sigma_{F_{[3.6]}}$ & $F_{[4.5]}$ & $\sigma_{F_{[4.5]}}$ & Telescope \\
 & & day & mJy & mJy& mJy& mJy& mJy& mJy& mJy& mJy& mJy& mJy & \\ \hline
2016-05-30 & 57538.5 & -345 & --- & --- & --- & --- & 0.030 & 0.003 & --- & --- & --- & --- & Keck/MOSFIRE \\
2016-10-11 & 57672.1 & -211 & 0.016 & 0.002 & 0.028 & 0.003 & 0.028 & 0.005 & --- & --- & --- & --- & P200/WIRC \\
2017-05-03 & 57876.5 & -7 & --- & --- & --- & --- & 0.04 & 0.01 & --- & --- & --- & --- & P200/WIRC \\
2017-05-09 & 57882.6 & -1 & --- & --- & --- & --- & 0.03 & 0.01 & --- & --- & --- & --- & Keck/MOSFIRE \\ \hline
2017-05-17 & 57890.0 & 7 & 28.92 & 0.53 & 21.93 & 0.40 & 14.80 & 0.27 & --- & --- & --- & --- & AZT/SWIRCAM (a) \\
2017-06-10 & 57914.5 & 31 & 33.82 & 4.67 & 22.34 & 1.65 & 17.95 & 1.65 & --- & --- & --- & --- & P200/WIRC \\
2017-06-16 & 57920.0 & 37 & 36.07 & 1.33 & 28.38 & 1.31 & 21.99 & 0.81 & --- & --- & --- & --- & MIRO (b) \\
2017-09-13 & 58009.7 & 126 & --- & --- & --- & --- & --- & --- & 4.441 & 0.003 & 6.132 & 0.006 & Spitzer/IRAC \\
2017-10-03 & 58029.1 & 146 & --- & --- & --- & --- & 4.81 & 0.44 & --- & --- & --- & --- & P200/WIRC \\
2017-10-10 & 58036.2 & 153 & --- & --- & --- & --- & --- & --- & --- & --- & --- & --- & P200/WIRC \\
2017-10-30 & 58056.0 & 173 & 4.80 & 0.27 & 3.67 & 0.27 & 2.30 & 0.21 & --- & --- & --- & --- & MIRO (b) \\
2017-11-23 & 58080.8 & 197 & --- & --- & --- & --- & --- & --- & 1.493 & 0.002 & 4.133 & 0.001 & Spitzer/IRAC \\
2017-11-25 & 58082.1 & 199 & --- & --- & 2.66 & 0.17 & 2.08 & 0.15 & --- & --- & --- & --- & P200/WIRC \\
2017-12-05 & 58092.0 & 209 & 3.71 & 0.24 & 2.52 & 0.23 & 1.64 & 0.18 & --- & --- & --- & --- & MIRO (b) \\
2018-01-11 & 58129.4 & 246 & --- & --- & --- & --- & --- & --- & 0.715 & 0.001 & 2.860 & 0.002 & Spitzer/IRAC \\
2018-02-25 & 58174.5 & 291 & 0.63 & 0.05 & --- & --- & 0.50 & 0.05 & --- & --- & --- & --- & P200/WIRC \\
2018-03-01 & 58178.2 & 295 & --- & --- & --- & --- & --- & --- & 0.418 & 0.001 & 1.796 & 0.001 & Spitzer/IRAC \\
2018-03-21 & 58198.1 & 315 & --- & --- & --- & --- & --- & --- & 0.336 & 0.002 & 1.521 & 0.001 & Spitzer/IRAC \\
2018-06-03 & 58272.6 & 389 & 0.29 & 0.03 & 0.30 & 0.05 & 0.18 & 0.01 & --- & --- & --- & --- & Keck/MOSFIRE \\
2018-06-23 & 58292.3 & 409 & 0.24 & 0.01 & --- & --- & 0.13 & 0.01 & --- & --- & --- & --- & P200/WIRC \\
2018-07-23 & 58322.3 & 439 & 0.18 & 0.01 & 0.19 & 0.01 & 0.08 & 0.01 & --- & --- & --- & --- & P200/WIRC \\
2018-09-12 & 58373.3 & 490 & 0.10 & 0.01 & 0.12 & 0.01 & 0.064 & 0.007 & --- & --- & --- & --- & P200/WIRC \\
2018-09-20 & 58381.9 & 498 & --- & --- & --- & --- & --- & --- & 0.095 & 0.001 & 0.316 & 0.001 & Spitzer/IRAC\\
2018-10-26 & 58417.3 & 534 & 0.057 & 0.005 & 0.071 & 0.006 & 0.046 & 0.005 & --- & --- & --- & --- & P200/WIRC \\
2018-11-27 & 58449.3 & 566 & --- & --- & --- & --- & --- & --- & 0.074 & 0.001 & 0.196 & 0.001 & Spitzer/IRAC \\
\\ \hline
\end{tabular}
\begin{scriptsize}
\tablenotetext{a}{\citealp{arkharov2017} using the Short-Wave Infrared Camera (SWIRCAM) on the AZT-24 telescope at Campo Imperatore Observatory.} 
\tablenotetext{b}{\citealp{rho2018} using Mount Abu Infrared Observatory (MIRO)} 
\end{scriptsize}
\end{table}

\subsection{Spectroscopy}\label{sec:sn_spec}
\begin{figure}
	\includegraphics[width=\linewidth]{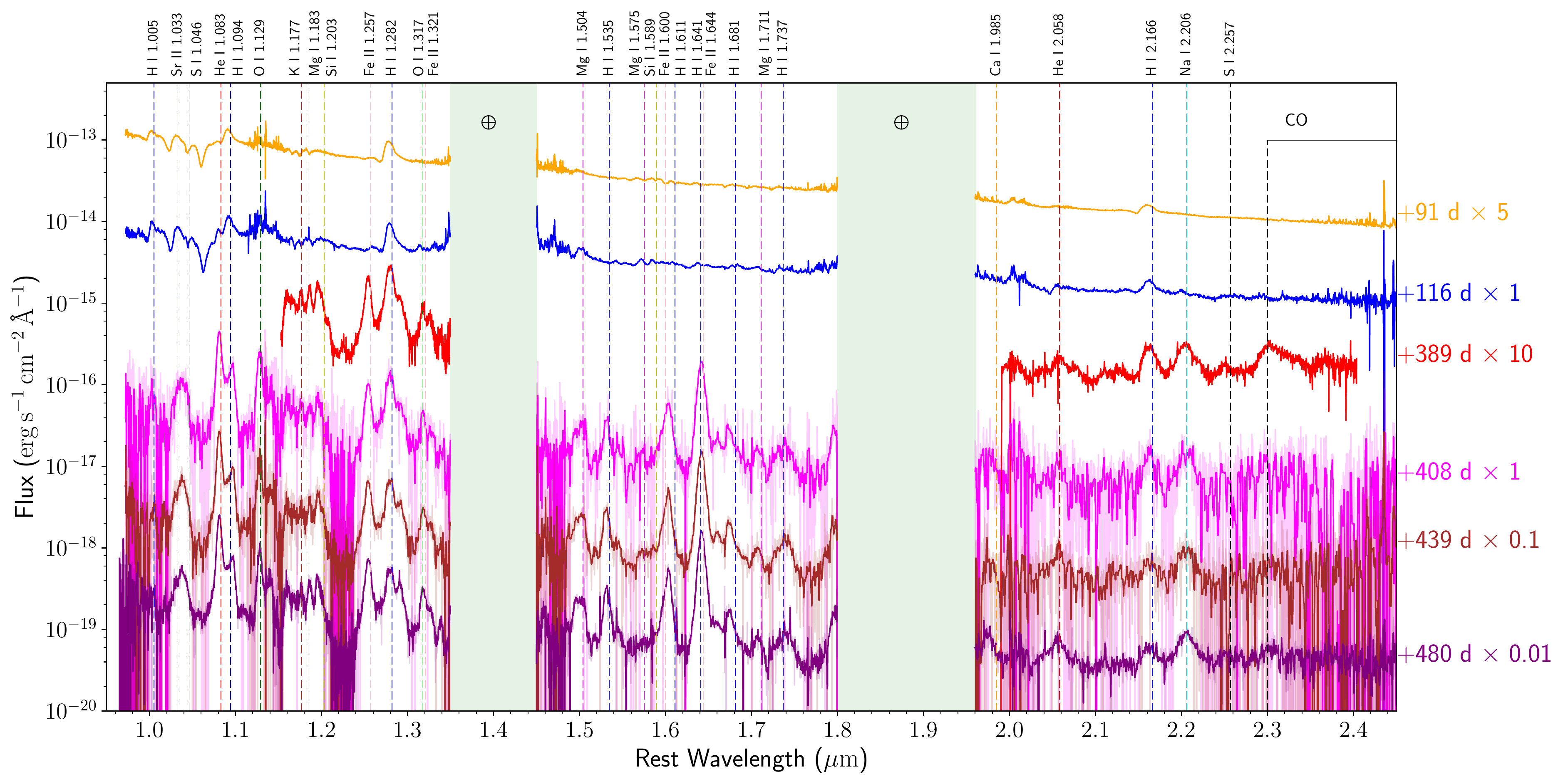}
	\caption{Near-IR spectra of SN\,2017eaw taken at 91, 116, 389, 408, 439, and 480 days post explosion. The wavelength plotted is corrected for the host galaxy's redshift of $z_{\rm host} = 0.00013$ \citep{epinat2008}. Rest wavelengths of common strong atomic lines are overplotted along with the CO vibrational first overtone band beyond 2.3 $\mu$m. The spectrum on +389 d was taken with Keck/MOSFIRE in the J and K bands only. Strong telluric bands around 1.4 and 1.85 $\mu$m are marked with $\oplus$ symbols. The flux at each epoch is multiplied by a factor indicated on the right for visualization. Spectra on 408, 439, and 480 d are smoothed by running median with 7, 7, and 3 pixel window respectively. Unsmoothed spectra are shown in transparent lines.}
    \label{fig:nir_specs}
\end{figure}

We obtained medium resolution (R$\sim$2,500) near-IR (1--2.5 $\mu$m; YJHK) spectroscopy with P200/TripleSpec \citep{herter2008} on 2017 Aug 9 (91 d), 2017 Sep 3 (116 d), 2018 Jun 22 (408 d), and 2018 Jul 23 (439 d).
TripleSpec is a long slit ($1"\times30"$) spectrograph, and we observed the SN using an ABBA dither pattern for sky subtraction. 
Type A0V standard stars HIP\,94140 and HIP\,75230 were observed either immediately before or after the SN observations to provide telluric correction and flux calibration.
Standard and SN observations were taken on different parts of the slit when the SN was observed after the standard to avoid persistence on the detector.
The reduction for \mbox{TripleSpec} data was done using a version of \texttt{Spextool} modified for TripleSpec \citep{cushing2004}. 
It applies field flattening, retrieves a wavelength solution from sky lines present in science observations, subtracts each AB pair to remove most of the sky emission, and then fits a low order polynomial to the different orders of spectral traces in the images. 
Simple spectral extraction is performed on the subtracted images (the optimal extraction algorithm \citep{horne1986} is not available in this version of the software0.
Telluric and flux calibrations were performed using \texttt{xtellcor} \citep{vacca2003}, which derives the instrument's efficiency by comparing the observed standard star spectrum with an A0V spectrum model from a high resolution spectrum of Vega. 
We obtained J and K band spectra with Keck/MOSFIRE on 2018 Jun 3 (389 d). 
Data reduction and spectral extraction were performed using MOSFIRE's data reduction pipeline\footnote{\url{https://keck-datareductionpipelines.github.io/MosfireDRP/}}, and telluric and flux calibrations were performed using \texttt{xtellcor}.
Finally, we obtained a 1--2.5 $\mu$m spectrum using the Near-Infrared Echellette Spectrometer (NIRES)\footnote{\url{https://www2.keck.hawaii.edu/inst/nires/}} on the Keck telescope on 2018 Sep 2 (480 d). 
The instrument is very similar to TripleSpec, except with a narrower 0.5" slit to take advantage of the better seeing from Maunakea. 
The observation strategy was similar.
The data were reduced using a version of \texttt{Spextool} modified for NIRES and the telluric and flux calibrations were performed using \texttt{xtellcor}.
The standard spectra used for this calibration was from HIP\,94140, observed before the SN. 
Fig. \ref{fig:nir_specs} displays our near-IR spectra for SN\,2017eaw from 91 to 480 d post-explosion with identifications of strong lines.
The identifications were guided by \cite{rho2018} and \cite{meikle1993}.
Each epoch is multiplied by a factor listed on the right of the figure for visualization. 
TripleSpec spectra at 408 and 439 d are smoothed by a running median with a 7 pixel window and the NIRES spectrum at 480 d is smoothed in the same way with a 3 pixel window. 
The unsmoothed version of the spectra are plotted in transparent lines. 

\section{Analysis and discussion}\label{sec:analysis}
\subsection{Progenitor non-variability and spectral energy distribution}\label{sec:prog_sed}
Some SNe II-P have shown stronger signs of CSM interaction than that of a typical SN II-P. 
This is indicative of a denser CSM in comparison to what expected from steady-state RSG wind driven mass loss.
The required enhanced mass loss would result in a variability in the progenitor's light curve, which may be a gradual brightening over timescales of years, or short term variability in case of eruptive mass loss.  
Such nearby and recently ejected CSM has been inferred in a variety of ways for a number of SNe II-P. 
Some SNe show signs of early dust formation and intermediate velocity (few$\times 10^3 \, \rm km\,s^{-1}$) hydrogen emission lines and/or X-Ray from the shocked CSM gas, e.g. SNe\,2007od \citep{andrews2010}, 2011ja \citep{Chakraborti2013, andrews2016}, and 2013ej \citep{Chakraborti2016,Mauerhan2017}. 
Some, like SN\,2009kf, show early strong ultraviolet emissions that either require a very energetic explosion or some degree of CSM interaction \citep{botticella2010,moriya2011}. 
Lastly, an emerging class of SNe II-P with very early spectra show narrow emission lines coming from the nearby CSM being ionized by the UV flash of the shock breakout, e.g. SNe\,2013fs \citep{yaron2017} and 2016bkv \citep{Hosseinzadeh2018}.
These narrow lines disappear quickly (as opposed to SNe IIn with narrow emission lines near peak).

There are multiple proposed mechanisms for ejecting mass from a RSG within a decade pre-explosion, e.g. wave driven mass loss \citep{fuller2017,fuller2018} and heavy-element nuclear-burning instability (9-11 $M_\odot$ RSG, \citealp{woosley2015}).
In the wave driven scenario for RSGs, the predicted variability for the smallest eruption case is of order 20\%, ejecting $\sim$0.1 $M_\odot$ \cite[the lowest heating efficiency ($\eta = 1/3$) case in][]{fuller2017}. 
For the heavy-element burning instability scenario, \cite{smith_arnett2014} argued that a 9-11 $M_\odot$ RSG can exhibit a detectable flash when the silicon burning commences. 
Nevertheless, the variability of the RSG progenitor star itself, which accompanies the mass loss event, has yet to be directly observed for progenitors to SNe II-P.\footnote{Pre-SN eruption has been documented in some SNe Ibn/IIn, which have denser CSMs. The most well studied cases are SNe\,2006jc \citep[e.g.][]{foley2007, pastorello2008} and 2009ip \citep[e.g.][]{mauerhan2013}. The definitive class of progenitors to those strongly interacting SNe has yet to be identified.} 
In addition, and there are other mechanisms proposed to explain the early-time observations that do not involve any dense CSM surrounding the progenitor \citep[e.g.][]{kochanek2018}. 

Indeed, our progenitor photometry in the near-IR (Fig.~\ref{fig:progenitor_image} \textit{bottom left}) shows that SN\,2017eaw's progenitor is not variable in the Ks band from one year to one day pre explosion on a timescle greater than 200 days. 
To put an upper limit on the variability of the progenitor, we compute the uncertainty of the weighted mean flux from the progenitor's Ks band photometry presented in Table \ref{table:NIR_phot}.
The uncertainty of the mean is $0.25\sqrt{\Delta F_{K_s} = \Sigma_i(\sigma_{F_{K_s},i}^2)}$ where $i$'s are four epochs at 345, 211, 7, and 1 days pre-explosion. 
The weighted mean flux with the uncertainty is $\bar{F}_{K_s}  = 0.031 \pm 0.002 \, \rm mJy$. 
Assuming the distances of 7.72 Mpc to the SN, we obtain the variability limits of $\Delta \nu L_\nu \lesssim 6\times 10^{3} \,  L_{\odot}$.
The total luminosity is $\nu L_\nu = 8\times 10^{4} \,  L_{\odot}$, typical of a RSG.
This is consistent with the luminosity derived from fitting the SED, discussed later in this section.
This corresponds to the variability upper limit of 6\% over a year.
In comparison, the weakest variability presented in \cite{fuller2017} is of order 20\%.  
Further, short-term variability immediately before core-collapse is ruled out by our last 2 epochs of observations within a few days before the explosion. 
We note that our finding of Ks band non-variability does not conflict with that of \cite{kilpatrick2018}, which presented a 20\% increase in the 4.5 $\mu$m flux over 3 years pre-explosion, but no variability in the 3.6 $\mu$m band. 

The near-IR SED of the progenitor star in the near-IR is shown in Fig.~\ref{fig:progenitor_image} (\textit{right}). 
The photometry has been corrected for Galactic extinction \citep[$E(B-V) = 0.304$;][]{schlafly2011} with no additional host extinction. 
We fitted a blackbody curve to the SED and found the best-fit parameters to be $T = 2{,}500 \pm 200 \, \rm K$ and $L = 7 \pm 2 \times 10^4 \, L_{\odot}$. 
We note that the blackbody luminosity is consistent with $\nu L_\nu$ obtained from Ks band observations presented above. 
These parameters are consistent with the progenitor star of SN\,2017eaw being a RSG. 

The progenitor star of SN\,2017eaw has also been observed by the LBT search for failed SNe in the optical (UBVR bands) in 35 epochs from 9 years to a few months before the explosion \citep{johnson2018}. 
They found no significant stochastic variability in the luminosity down to $\Delta \nu L_\nu \lesssim$ 700 $\rm L_{\odot}$ in V and R bands. 
\cite{johnson2018} also reported non-variability in other 3 progenitor stars to SNe\,2013am, SN 2013ej, and ASASSN-2016fq.
One of these SNe, SN\,2013ej, has a nearby dense CSM inferred from X-Ray \citep{Chakraborti2016} and optical spectropolarimetry \citep{Mauerhan2017} despite its progenitor having exhibited no variability in the last 5 years before the explosion. 
They argued that while outbursts may happen on a timescale shorter than their observing cadence, it is unlikely that the decline from these outbursts is fast enough to escape a detection because the dynamical and thermal timescales for these RSGs are much longer ($\sim$years) than the observational cadence ($\sim$months).

Our observations, along with those of \cite{johnson2018}, do not find major eruptive mass loss events around SN\,2017eaw's progenitor in the last 10 years of its life. 
The CSM around SN\,2017eaw, inferred from radio and X-Ray observations, \citep{nayana2017b, kong2017, grefensetette2017}, may be ejected in minor mass loss events which cause variability smaller than 6\% or last shorter than our observational cadence. 
Alternatively, the CSM around the RSG progenitor may be more akin to the compact CSM shell observed around Betelgeuse \citep{bertre2012}. 
\citet{mackey2014} presented a scenario in which such a compact CSM shell is constructed by a progenitor wind being trapped by ionizing photons in the star's environment.
The CSM shell will eventually interact with the SN shock, and as shown by \cite{smith2009}, this interaction will not be strong enough to produce typical Type IIn narrow lines.
However, weak signs of CSM interactions may be detected in the X-ray and radio. 
For comparison, \cite{pooley2002} derived from 0.2-10 keV X-ray observations that SN\,1999em (typical II-P) has a progenitor mass loss rate of $\dot{M} \sim 2 \times 10 ^{-6} \, M_{\odot}\, \rm yr^{-1}$, similar to that of Betelgeuse. 
From the literature, the 0.3-10 keV X-ray luminosity of SN\,2017eaw is $1.1 \times 10^{39}\, \rm erg\, s^{-1}$ at 11 d post explosion \citep{grefensetette2017}. 
In comparison, SN\,1999em's X-ray luminosity is $2 \times 10^{38} \, \rm erg\, s^{-1}$ at 4 d post explosion, \citep{pooley2002}.
Since the X-ray luminosity scales with $\dot{M}^2$ (see equation (3.10) from \citealp{fransson1996}), SN\,2017eaw's progenitor would have a mass loss rate of $\dot{M} \sim 5 \times 10 ^{-6} \, M_{\odot}\, \rm yr^{-1}$, which is in the typical range for a RSG.
This mass loss rate is consistent with the figure of $9\times10^{-7} \, M_{\odot} \, \rm yr^{-1}$ derived by \cite{kilpatrick2018} using \textit{Hubble} and \textit{Spitzer} photometry.
Hence, it is possible that the CSM for SN\,2017eaw is created by a trapped wind like that seen around Betelgeuse, and not by eruptive mass loss in the last decade of the progenitor's life.

\subsection{Photometric evolution and comparison}\label{sec:LC_spec_comparison}
In order to assess SN\,2017eaw's place in the Type II-P population, we compare its IR photometric evolution to those of other well studied SNe II-P.
We first caution that SN evolution in the near-IR remains poorly sampled in the nebular phase, making direct comparison to other individual SNe difficult.
In the optical, \cite{tsvetkov2018} has shown that SN\,2017eaw bears photometric similarities to SN\,2004et both in terms of flux and color evolutions.
Fig.~\ref{fig:compare_17eaw} \textit{Top} shows SN\,2017eaw's 1-5 $\mu$m light curves in comparison to those of SN\,2004et \citep{kotak2009,maguire2010,fabbri2011} and SN\,2002hh \citep{pozzo2006}, two normal SNe II-P in NGC\,6946. 
SN\,2017eaw is fainter than SN\,2004et by about 0.5 mag in all bands, including the optical \citep{tsvetkov2018}, without a strong wavelength dependence, which indicates that SN\,2017eaw is intrinsically less energetic and not that it suffers more extinction and reddening. 
In \textit{Spitzer} 3.6 and 4.5 $\mu$m bands, the evolutions of SNe 2017eaw and 2004et are very similar. 
Except for the 4.5 $\mu$m band, other near-IR bands show a similar linear decline of $\sim$1.5 mag/100d after 120~d. 
The 4.5 $\mu$m band, however, declines at a slower rate of 0.80 mag/100d.
For SN\,2004et, \cite{kotak2009} showed, using a series of SEDs with all \textit{Spitzer} bands from 3.6 to 24 $\mu$m, that this different decline rate in the 4.5~$\mu$m band is due to the CO fundamental vibrational emission at 4.65~$\mu$m.
From this comparison alone, we can infer that CO is forming in SN\,2017eaw.
This is consistent with the emerging CO first overtone emission band starting around 120 d reported by \cite{rho2018}. 
Our spectra, presented above in \textsection\ref{sec:sn_spec} and discussed in \textsection\ref{sec:co_evo}, also confirm the presence of CO.
After 250 d, the decline rate in the J, H, and Ks bands are 1 mag/100d, the canonical decline rate for a light curve powered by $\rm ^{56}Co$ decay. 


Fig.\ref{fig:compare_17eaw} \textit{Bottom} shows SN\,2017eaw's \textit{Spitzer} light curves in comparison to all other SNe II-P light curves observed by \textit{Spitzer}, aggregated by \cite{szalai2018}. 
Data here include photometry from \cite{tinyanont2016}. 
SN\,2017eaw falls in the middle of the distribution of SNe II-P absolute magnitudes in both the 3.6 and 4.5 $\mu$m bands, suggesting that it is a typical SN II-P.
That SN\,2017eaw is typical implies that results we derive  for SN\,2017eaw in this paper may be more generally applicable to other SNe II-P that are fainter and more difficult to observe. 
More specifically, SN\,2017eaw's (and SN\,2004et's) decline rates of 1.5 and 0.8 mag/100d in the 3.6 and 4.5 $\mu$m bands are typical among other SNe II-P with observations between 100-300 d. 
This suggests that strong CO emission in the 4.5 $\mu$m band may be ubiquitous among SNe II-P. 
The implication here is that previous studies of SN dust based on warm \textit{Spitzer} data with only 3.6 and 4.5 $\mu$m data may provide unreliable dust estimates since the 4.5 $\mu$m band is dominated not by thermal emission from dust, but CO line emission. 
The result is that the dust luminosity---and, consequently, dust mass---is overestimated. 

While SN\,2017eaw's early temporal evolution is similar to that of most other SNe II-P, we note that it is markedly different from SNe\,2004dj \citep{kotak2005} and 2011ja \citep{andrews2016,tinyanont2016}.
Both of these SNe show signs of CSM interaction and dust formation at much earlier epochs: 65-165 d for SN\,2004dj \citep{meikle2011} and 105 d for SN\,2011ja \citep{andrews2016}.
Lastly, SN\,2017eaw's resemblance to SN\,2004et's \textit{Spitzer} light curve presents an intriguing possibility that SN\,2017eaw will rebrighten just like SN\,2004et at $\sim$1,000 d due to shock interaction with a distant CSM shell (see Fig. \ref{fig:compare_17eaw} \textit{bottom} and \citealp{kotak2009}).
This possibility warrants continued monitoring of this SN, and other nearby CCSNe in the future, in the IR either by \textit{Spitzer} or ground based instruments.
More generally, future IR observations of nearby CCSNe will reveal a range of epochs when CSM interactions and/or dust formation commence, from $\sim$100 to $\sim$1,000 d post explosion. 
Such observations will provide clues to the physical processes in the last stage of RSG evolution that are responsible for the distant CSM shell seen in e.g. SN\,2004et. 

\begin{table}
\centering
\scriptsize
\caption{Properties of SNe used to compare with SN\,2017eaw}
\label{table:compare_sne}
\begin{tabular}{lllp{2.5in}p{2in}} \toprule
Name & Type & Host & Comparison points & References \\ \hline
2017eaw & II-P & NGC\,6946 & & \cite{rho2018, tsvetkov2018} \\ \hline
1987A & II-pec & LMC & CO evolution and line profile & e.g. \cite{spyromilio1988, liu1992, meikle1989, meikle1993}\\
1999em & II-P & NGC\,1637 & Typical SN II-P X-ray luminosity & \cite{pooley2002} \\
2002hh & II-P & NGC\,6946 & Photometric similarities in optical and IR & \cite{pozzo2006}  \\
2004et & II-P & NGC\,6946 & Photometric similarities in optical and IR & \cite{maguire2010} \\ 
       &      &           & Silicate dust detection &        \cite{kotak2009}                \\ 
2004dj & II-P & NGC\,2403 & \textit{Spitzer} light curves for SNe II-P with early CSM interactions & \cite{kotak2005, meikle2011}\\
2011ja & II-P & NGC\,4945 & \textit{Spitzer} light curves for SNe II-P with early CSM interactions & \cite{andrews2016, tinyanont2016} \\
\hline
\end{tabular}
\end{table}

\subsection{Spectral energy distribution modeling}\label{sec:sed}
We performed radiative transfer (RT) modeling of the SEDs for SN 2017eaw to estimate the mass of dust associated with the SN. We used MOnte CArlo SimulationS of Ionized Nebulae (MOCASSIN; version 2.02.72), which is a fully self--consistent 3D Cartesian dust RT code \citep{2003MNRAS.340.1136E, 2005MNRAS.362.1038E, 2008ApJS..175..534E}. 
The models accept multiple user inputs, but due to the limited wavelength coverage in our SED data, we limited the number of parameters we fitted, in order to avoid overfitting. 
While MOCASSIN takes a series of user inputs, such as number of dimensions, grid size, dust density, composition, and distribution, the only parameters we varied to fit the SED were the inner and outer radii of the dust shell, and the density at the inner edge of the shell. 
We discuss the other parameters, which are held constant, later in this section. 
The luminosity and temperature of the photon source (the SN) were measured or extrapolated from the optical data. 
Any interactions, whether absorption or scattering, between photons and dust grains are governed by Mie scattering theory. 
MOCASSIN returns the temperature, mass, and opacity of the dust shells. 

We used MOCASSIN to calculate the mass of dust at 126, 197, 295, and 498 d, epochs for which we had \textit{Spitzer} coverage.
The optical data for 126 and 197 d are from \cite{tsvetkov2018}, while the 295 d epoch is extrapolated (plotted as open symbols).
We checked that the extrapolated photometry did not differ significantly from SN\,2004et photometry at the same epoch.
We did not extrapolate to 498 d, and simply used the same blackbody input for the 295 and 498 d epochs. 
The near-IR data for 126 d are from the flux calibrated TripleSpec spectrum. 
Optical through near-IR photometry have been corrected for Galactic extinction ($E(B-V) = 0.304$; \citealp{schlafly2011}).
We assumed, according to \cite{tomasella2017}, that the host galaxy extinction is negligible.

We chose to model this system as a central point source surrounded by a gas--free dust shell. 
The shell was further assumed to be ``smooth'', which means that there are no inhomogeneities (``clumps''), with the dust density profile falling by r$^{-2}$ from the inner radius (R$_{\rm in}$) to the outer radius (R$_{\rm out}$). 
We used the standard Mathis Rumpl Nordsieck \citep[MRN;][]{1977ApJ...217..425M} power law distribution, ${\rm a}^{-3.5}$, to specify the size distribution of the dust grains. 
We tested two compositions for the dust grains: 100$\%$ amorphous carbon (amC) and 100$\%$ silicate grains \citep[respectively]{1988ioch.rept...22H,1992A&A...261..567O}. 
We were unable to constrain the composition due to a lack of data beyond 4.5 $\mu$m.  
We do not fit the R, I, and 4.5 $\mu$m bands in our SEDs because they are contaminated by strong emission from H$\alpha$, the IR Ca II triplet, and CO bands, respectively. 
Our fits are not unique. 
We used the shape of the SED to set our input parameters. 
We used an inner radius R$_{\rm in}$ of $10^{16}$ cm and an outer radius R$_{\rm out}$ of $10^{17}$ cm as initial inputs, based on the optical and IR contributions to the SED. 
The input optical luminosity and temperature were based on the optical continuum.

\begin{figure}
\centering
	\includegraphics[width = 4in]{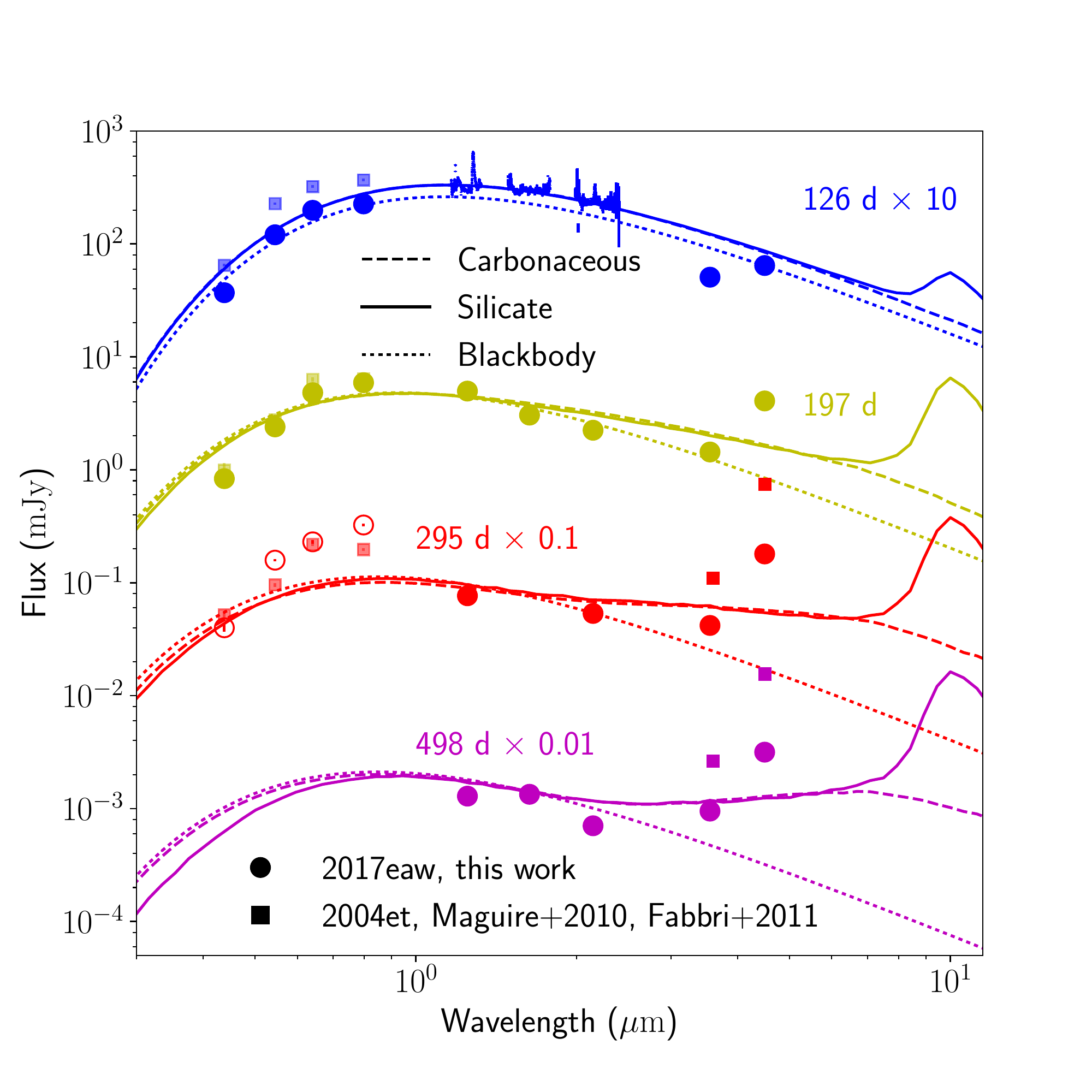}
    \caption{Evolution of the SED of SN\,2017eaw (circle points, optical data from \citealp{tsvetkov2018}) in comparison to SN\,2004et (square points, from \citealp{maguire2010}) at 126, 197, 295, and 498 d. 
    Optical data for the 295 d epoch are extrapolated, and are shown with open symbols. 
    SED models of blackbody with dust grain emission from MOCASSIN are overplotted. Dashed lines represent model with purely carbonaceous dust grains while solid lines represent those with purely silicate grains. Dotted lines are blackbody fits, provided for comparison.
    The fits ignore optical R, I bands, and \textit{Spitzer} 4.5 $\mu$m band due to contaminations from H$\rm \alpha$, Ca triplet, and CO fundamental band respectively.}
    \label{fig:SED}
\end{figure}

At 126 days, the input SED was our best fit, indicating no dust was detected. 
Starting from 197 but more clearly at 295 and 498 d, a small amount of dust ($\sim 7 \times 10^{-6}$ M$_{\sun}$ for carbon grains and $\sim 10^{-4}$  M$_{\sun}$ for silicate grains) was required to fit the SED. 
The peak temperature of the dust is $\sim$500 K.
Fig. \ref{fig:SED} shows the SED evolution of SN\,2017eaw, compared with SN\,2004et \citep{maguire2010,fabbri2011}, with the best-fit models from MOCASSIN overplotted for each epoch. 
Solid and dashed lines represent amorphous carbon and silicate grains respectively.
Dotted lines are blackbody provided for comparison. 
The excess flux in the 4.5 $\rm \mu m$ band is similar in both SNe.
For SN\,2004et, it was shown to be due to emission from the CO fundamental band \citep{kotak2009}.
Our data cannot distinguish between carbonaceous and silicate dust due to the lack of data around 10 $\mu$m.
However, we note that in the case of SN\,2004et, a broad silicate and SiO feature around 10 $\mu$m were detected, pointing to some silicate grains \citep{kotak2009}.
Furthermore, dust condensation models predict that for SNe II-P, carbonaceous dust does not form until $\sim$1,000 d post-explosion (discussed above) while silicate dust (like forsterite) can start to form as early as 200 d \citep[Fig. 5 in][]{sarangi2015}. 
The comparison with SN\,2004et and models suggest that dust responsible for SN\,2017eaw's 3.6 $\mu$m excess at 295 d may be $\sim 10^{-4}$  M$_{\sun}$ of silicates. 
In summary, the SED evolution of SN\,2017eaw is similar to that of SN\,2004et.
A single component blackbody can fit the SED reasonably well for the 126 and 197 d epochs while a small amount of dust, likely $\sim 10^{-4}$  M$_{\sun}$ of silicate dust is required to fit the 295 and 498~d epochs. 
Finally we note that the rising IR excess is in agreement with \cite{rho2018}'s report of rising flux in the red part of their K band spectra starting at $\sim$120~d. 



\subsection{Spectroscopic evolution of the CO first overtone feature}\label{sec:co_evo}
Spectra of SN\,2017eaw shown in Fig. \ref{fig:nir_specs} evolved from the photospheric phase to the nebular phase after the 116 d epoch.
The photospheric spectra at 91 and 116 d were dominated by strong continuum with hydrogen lines with P-Cygni profiles. 
Metal lines started to emerge from 116 d to 389 d as the SN transitioned into the nebular phase.
The helium line at 1.083 $\mu$m emerged at 116~d and became strong in the nebular phase, blended with Pa-$\gamma$. 
Numerous metal species were also present in the nebular spectra, notably [Fe II] at 1.257, 1.600, and 1.644 $\mu$m. 
The latter two were blended with hydrogen. 
Na I at 2.206 $\mu$m emerged at 116 d and became as strong as Br-$\gamma$ in the last 4 epochs. 
Modeling these spectra to obtain physical parameters for the ejecta are outside the scope of this paper. 

Our series of near-IR spectra capture the evolution of the 2.3-2.5 $\mu$m band of the CO vibrational first overtone ($\Delta v = 2$) transition.
\cite{rho2018} reported that the feature emerged at 124 d and strengthened until their last epoch at 205 d. 
Assuming a LTE level population, they inferred a growing CO mass from 0.6-1.6$\times 10^{-4} M_\odot$ at 124~d to 1.9-2.2$\times 10^{-4} M_\odot$ at 205 d, with the temperature declining in their last 3 epochs, reaching 2,700~K at 205~d. 
Our spectrum at 389 d showed that the CO feature continued to strengthen and had become as prominent as hydrogen by this epoch. 
Spectra from 408, 439, and 480~d subsequently showed that the CO feature started to fade. 
The series of these last four spectra are shown in Fig. \ref{fig:CO_line} \textit{left} in the linear scale to demonstrate the fading CO feature.
Wavelengths for CO band heads of the $\Delta v = 2$ transitions are overplotted, both for $\rm ^{12}CO$ and $\rm ^{13}CO$ with $v = -500 \rm \,km\,s^{-1}$. 

The fading of the CO first overtone feature may indicate either that CO molecules have been destroyed or that the gas has cooled down enough that the $v \geq 2$ vibrational levels are no longer excited. 
Some authors argued that CO can be easily destroyed by high energy electrons from $\rm ^{56}Co$ decay and that most atomic carbon condenses into amorphous carbonaceous dust grains starting at 200-300 d post explosion \citep{todini2001}. 
However, in this scenario, the CO formation would have been inhibited in the first place (not formed and destroyed later) and their model predicted the CO mass for SN\,1987A to be a factor of 3 smaller than what was observed.
They did not address how dust grains could survive the same energetic electrons from $\rm ^{56}Co$ decay that destroyed the CO molecules.

Alternatively, the fading of the first overtone feature may indicate that the ejecta where CO formed have cooled enough that the $v \geq 2$ vibrational levels corresponding to the first overtone emission are no longer excited.
The scenario is further supported by the fact that the 4.5 $\mu$m excess in the SED (see \textsection\ref{sec:sed}), likely due to the CO fundamental emission, did not disappear.
This is because the fundamental transition can happen as long as the $v = 1$ level, which is less energetic, remains populated. 
Another line of evidence that CO did not get destroyed in SN\,2017eaw's ejecta is the similarity between its CO line evolution and that of SN\,1987A. 
Recall that in SN\,2017eaw, the first overtone bands emerged at $\sim$200 d and started to disappear at $\sim$400 d while the fundamental bands (4.5 $\mu$m band) remained detected at 566 d. 
This behavior is consistent with what was observed in SN\,1987A where the CO first overtone emerged as early as 100 d and faded by 574 d while the CO fundamental band remained strong until 600-700 d \citep[e.g.][]{spyromilio1988, meikle1989, liu1992, meikle1993}.
The survival of CO in SN ejecta was shown for SN\,1987A by late time observations with ALMA, more than 25 years post explosion, that detected CO (along with SiO) rotational emission \citep{kamenetzky2013, abellan2017}.
In summary, the fading CO first overtone feature indicates that the ejecta have cooled, and not that CO molecules are destroyed. 

The survival of CO is favored by more recent chemical evolution models which treat chemical reactions in SN ejecta more realistically \citep[Molecular Nucleation Theory, e.g.][]{sarangi2013,sarangi2015, sluder2018}.
Specifically, the chemical evolution of the SN ejecta is modelled using a realistic network of chemical reactions to track molecules and dust formation together.
\cite{sarangi2013, sarangi2015}, assuming stratified ejecta, found that in the ejecta layers where carbon is abundant (zones 4 and 5 in their papers), CO forms in the oxygen-rich layer (zone~4) and carbonaceous dust can only form in the oxygen-poor part (zone 5). 
This is because CO is more easily formed and is not destroyed once formed, so that all carbon atoms in zone 4 end up in CO. 
\cite{sluder2018} used a realistic ejecta model that has some radioactive species mixed into the ejecta to model molecule and dust dissociation due to the high energy electrons from radioactive decays. 
They concluded that, while high energy electrons keep the ejecta gas from becoming fully molecular, they do not destroy all CO, and that up to 0.06 $M_\odot$ of CO survives to 10,000 d (see their Figure 14). 
The implication of this result is that CO is not easily destroyed to form carbonaceous dust as was concluded by \cite{todini2001}. 

\subsection{Line profile of the CO first overtone feature}\label{sec:co_profile} 
\begin{figure}
	\centering
	\includegraphics[width = \textwidth]{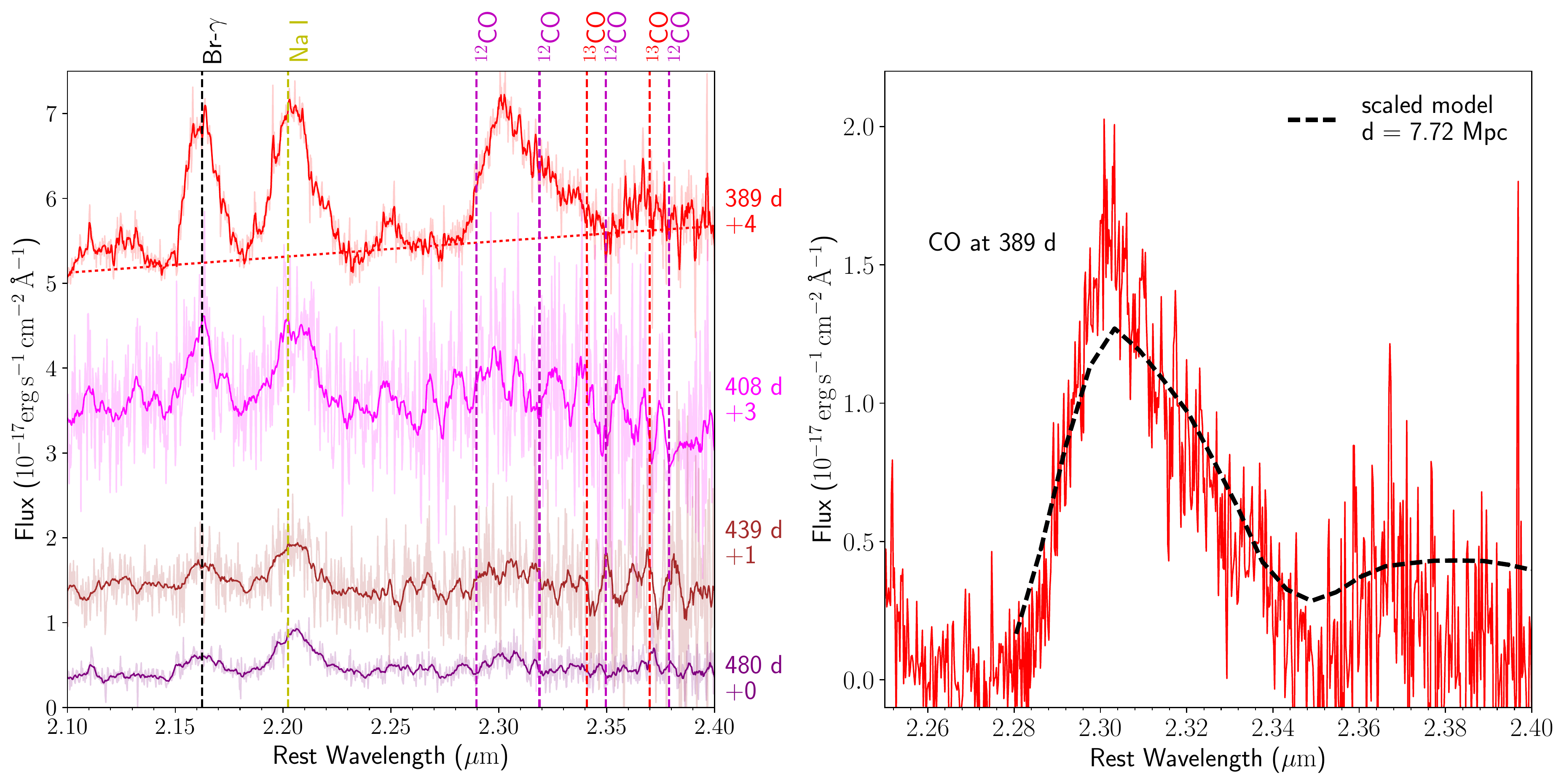}
    \caption{\textit{Left} The evolution of K band spectra of SN\,2017eaw from 389 to 480 d. The colors correspond to the same epochs as in Fig. \ref{fig:nir_specs}. Text on the right of the spectrum from each epoch indicates the epoch and the offset applied for visualization. The window size for smoothing these spectra are 5, 21, 21, and 17 pixels respectively for visualization and the unsmoothed spectra are shown in transparent lines. Wavelengths of the Br-$\gamma$, Na I, and CO bandheads, all with $v = -500 \rm \, km\,s^{-1}$, are plotted. The red dotted line is the blackbody fit to the continuum at 389 d with temperature and blackbody radius $T = 804\,\rm K$ and $r_{BB} = 4\times10^{15}\,\rm cm$. \textit{Right} CO line profile of SN\,2017eaw at 389 d with the fitted continuum shown on the left subtracted. Dashed lines are the non-LTE model fitted to the spectrum of SN\,1987A at 377 d post explosion \citep{liu1992}. The model has been shifted with $v = -500 \, \rm km\,s^{-1}$, which is estimated from the Br-$\gamma$ line from the same epoch. } 
    \label{fig:CO_line}
\end{figure}

The CO emission from SN\,2017eaw is similar to that from SN\,1987A, not only in its temporal evolution, but also in the line profile. 
In this subsection, we consider our 389 d epoch, which has the strongest detection of the CO feature. 
To isolate SN\,2017eaw's CO line profile, we first estimated and subtracted the underlying continuum by fitting a blackbody curve to parts of the K band spectrum at 389 d without strong lines.
The blackbody temperature and radius are $T = 809\, \rm K$ and $r_{\rm BB} = 4\times10^{15}\,\rm cm$ respectively.
The model is shown in Fig.~\ref{fig:CO_line} \textit{left}. 
This warm continuum is likely coming from the same component that \cite{rho2018} reported in their spectra.
By 480 d, this continuum has substantially decreased. 
The continuum subtracted CO profile is shown in Fig.~\ref{fig:CO_line} \textit{right}. 

The CO line profile of SN\,2017eaw observed on day 389 is very similar to that of SN\,1987A at 377~d \citep{spyromilio1988}.
On top of the continuum subtracted CO line profile shown in Fig. \ref{fig:CO_line} \textit{right}, we plotted the non-LTE model by \cite{liu1992}, which was fitted to SN\,1987A's spectrum at 377 d.
We shifted the wavelength with $v = -500 \, \rm km\,s^{-1}$ (blueshift), a line peak velocity estimated from the Br-$\gamma$ line from the same epoch. 
We only scaled the flux of the model by the squared ratio of the distances to SNe\,1987A (51.4 kpc, \citealp{panagia2003}) and 2017eaw.
The result assuming $d=$ 7.72 Mpc fitted our data reasonably well.
We next describe the \cite{liu1992} model and how we can use it to explain SN\,2017eaw. 

The \cite{liu1992} model predicts the CO emission features while accounting for non-LTE effects by assuming that the CO vibrational level populations are determined by collisional excitation and radiative de-excitation. 
This is a reasonable assumption because, at low temperature and density at this epoch, the collisional timescale is an order of magnitude longer than the radiative timescale, rendering collisional de-excitation ineffective. 
\cite{liu1992} contrasted their non-LTE results with the LTE results from \cite{spyromilio1988} to demonstrate that the level population at this epoch is clearly non-LTE as their model provided a superior fit to the data (see their figures 1 and 2 for LTE and non-LTE results respectively).
Further, the non-LTE models implied an order of magnitude higher CO mass for SN\,1987A in comparison to LTE models.
For the particular model of SN\,1987A we use for comparison with to our SN\,2017eaw data, the CO mass is $9.1\times10^{-4}\,  M_{\odot}$, the temperature is 1,800~K, and the ejecta velocity is 2,000 $\rm km\,s^{-1}$. 

The fact that the model fitted to SN\,1987A data at 377 d explains our SN\,2017eaw data well suggests that the CO properties (mass, temperature, and velocity) between the two SNe are similar. 
This implied CO mass of $\sim 10^{-3} \,  M_{\odot}$ for SN\,2017eaw is a factor of a few higher than the values determined by \cite{rho2018}; however we note earlier that this discrepancy is expected from comparing non-LTE to LTE models. 
%
The line ratios between different transitions in this band is determined by the relative population in different vibrational excitation states. 
This is determined by the electron temperature since CO is excited by collisions with free electrons in this model \citep{liu1992}.
Comparing SN\,2017eaw's CO profile to different models at $T$ = 1,800 - 4,000 K from \cite{liu1992}, it is clear that the plotted model at 1,800 K fits our data best. 
Finally, the width of the profile is determined by the ejecta velocity, and again, this model with $v = 2{,}000\, \rm km\,s^{-1}$ explains our data well.
In summary, we are able to estimate electron temperature, CO mass, and velocity for SN\,2017eaw at 389~d by comparing its observed CO first overtone line profile with a non-LTE model for SN\,1987A's CO at 377~d.
The similarity in line profiles indicates that the two SNe have similar CO mass, temperature, and velocity at this epoch, which is surprising since SN\,1987A is a very different kind of explosion coming from a different type of progenitor (blue supergiant (BSG), instead of RSG). 
Future work fitting the entire sequence of SN\,2017eaw spectra with non-LTE models is required to get a complete picture of its CO mass evolution. 


\subsection{Molecule and dust formation in comparison to chemical evolution models}\label{sec:molecule_dust}
Near-IR spectroscopy, along with cadenced \textit{Spitzer} observations, provided a detailed monitoring of the chemical evolution of SN\,2017eaw's ejecta, which can be compared to different chemical evolution models.  
The series of near-IR spectra presented by \cite{rho2018} and this work are the best nebular phase near-IR spectra of any SNe since SN\,1987A. 
From previous subsections, we have presented the following.
(i) CO formed in SN\,2017eaw's ejecta by $\sim$200 d (observed by \citealp{rho2018}) and cooled such that the first overtone transitions were no longer excited by $\sim$400~d. The 4.5~$\mu$m band excess, likely due to the less energetic fundamental transitions, was still detected at 566~d, showing that the CO molecules were not destroyed. 
(ii) A contribution of hot dust emission to the 3.6~$\mu$m channel has been detected starting at 300~d post explosion, indicating the presence of warm dust (which may be newly formed or pre-existing).
We estimated the dust mass of $7 \times 10^{-6} \, M_\odot$ and $10^{-4} \, M_\odot$ assuming pure carbonaceous and silicate grains respectively. 
We now compare these observations with predictions from the simpler classical nucleation theory along with those from the more realistic molecular nucleation theory.  

First, we consider the formation and survival of CO. 
CNT models generally consider molecule and dust formations as different processes.
The only influence molecules have on dust formation is that they deplete different species in the gas phase available to form dust. 
For example, \cite{todini2001} show, assuming a complete mixing of radioactive species throughout the ejecta, that most carbon atoms end up in amorphous carbonaceous dust grains as CO molecules are destroyed by energetic electrons from $\rm ^{56}Co$ decays. 
MNT models, like \cite{sarangi2013, sarangi2015}, show that CO is not a precursor molecule to carbonaceous dust grain formation because CO and carbon dust form via different chemical pathways, in different parts of the ejecta. 
In their models, \cite{sarangi2015} assumed that the ejecta are stratified into different zones with different compositions. 
In the zones with abundant carbon, if oxygen is present, all carbon atoms are used up in CO.
Carbonaceous grains, on the other hand, only form in the oxygen poor part of the carbon rich ejecta. 
\cite{sluder2018} presented results from their MNT models using a more realistic ejecta model with some degree of mixing. 
They showed that up to $10^{-2}\, M_\odot$ of CO can form in the first 100~d post explosion in SN\,1987A-like ejecta (massive progenitor), and survive until 10,000~d where their simulation ends.  
The major role of CO in dust formation is that it radiatively cools the ejecta. 
SiO formation may as well play a role here, but we do not have direct evidence due to the lack of observations at wavelengths longer than $5\, \rm \mu m$ (though we note that SiO was observed in SN\,2004et). 
Our observations showing that a $\sim 10^{-3}\, M_{\odot}$ of CO have formed by 389 d, and that CO survive to at least 566 d agree with the scenario predicted by MNT. 

\begin{figure}
    \centering
    \includegraphics[width = \textwidth]{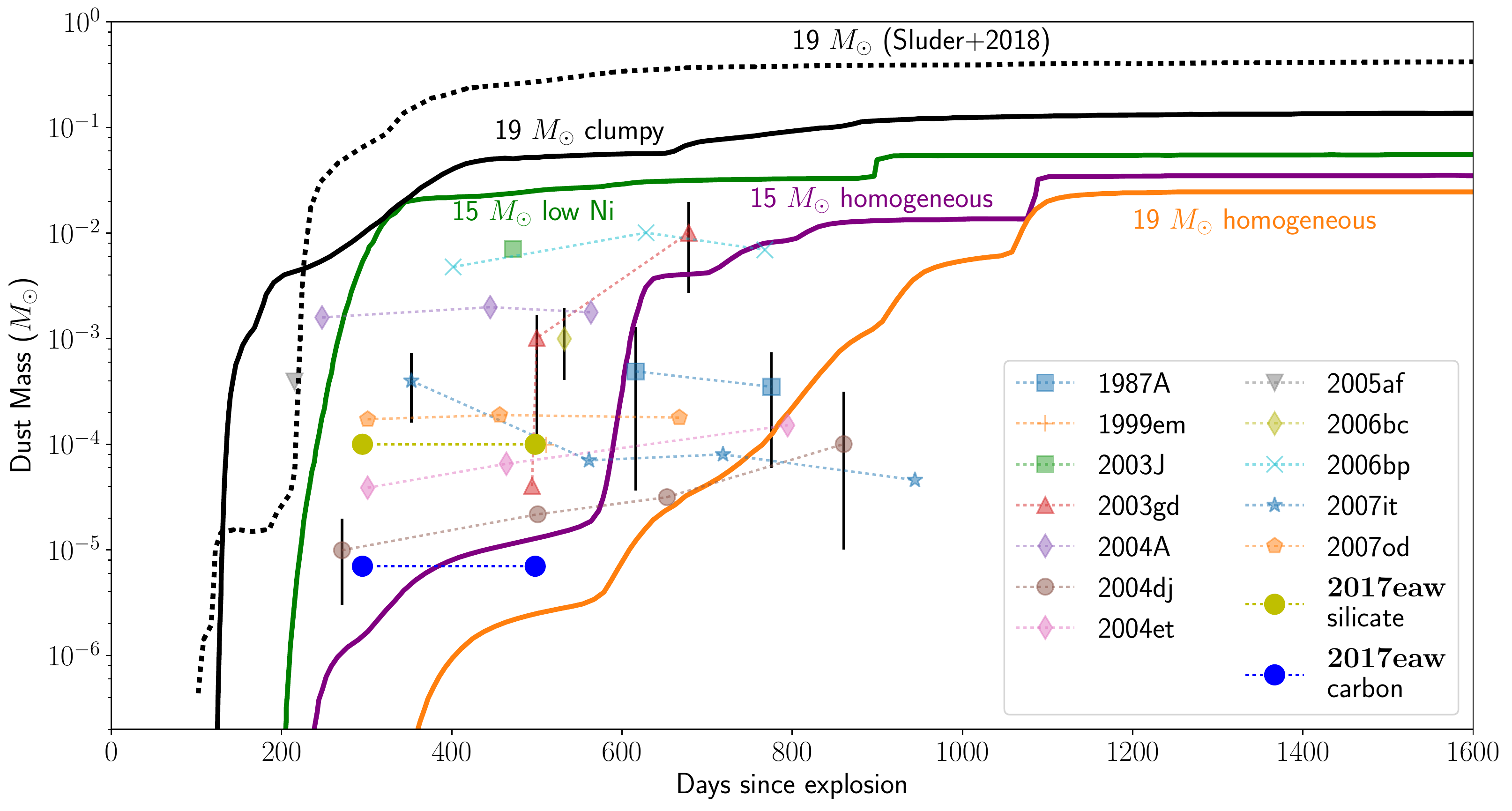}
    \caption{
    The dust mass evolution of SN\,2017eaw derived from the SED fitting in comparison to other SNe II-P and II-pec (1987A).
    Yellow and blue circles are results for 100\% silicate and 100\% carbon grains cases respectively.
    Other points are observed dust mass from other SNe from the literature, taken from Figure 10 of \cite{sarangi2015}. 
    Solid lines are dust mass predictions from different models in \cite{sarangi2015} with progenitor mass and ejecta assumptions notated.
    Dashed black line is the total dust mass from \cite{sluder2018} model specifically for SN\,1987A. 
    They noted that their model should be compared to the 19 $M_\odot$ with clumpy ejecta scenario from \cite{sarangi2015}. 
    We note that these models consider only dust formation in the SN ejecta and do not consider pre-existing dust and newly formed dust in the CSM. 
    }
    \label{fig:dust_mass_models}
\end{figure}

Secondly, we consider the formation of dust and the evolution of its mass.
Fig.~\ref{fig:dust_mass_models} shows the dust mass measurement of SN\,2017eaw in comparison to some other SNe and model predictions from \cite{todini2001}, \cite{sarangi2015}, and \cite{sluder2018}. 
CNT models tend to predict quick precipitation of dust in the SN ejecta, with \cite{todini2001} predicting 0.1 $M_\odot$ of amorphous carbon formed by 400 d and 0.4 $M_\odot$ of $\rm Mg_2SiO_4$ formed by 600 d. 
MNT, on the other hand, predicts a range of dust formation behavior depending on the progenitor mass and ejecta composition. 
For example, \cite{sarangi2015} predicted only $2\times10^{-2} \, M_\odot$ of $\rm Mg_2SiO_4$ at 300 d with carbonaceous dust not forming until 900 d in their 15 $M_\odot$ progenitor model with homogeneous SN ejecta and low (0.01 $M_\odot$) of $\rm ^{56}Ni$.
Their 15 $M_\odot$ progenitor model with a normal amount (0.075 $M_\odot$) of $\rm ^{56}Ni$ predicts a slower dust formation with the total dust mass reaching $2\times10^{-2} \, M_\odot$ at around 800 d. 
For their 19 $M_\odot$ models, they considered both homogeneous and clumpy ejecta.
They found that the dust condenses gradually in the homogeneous ejecta, only reaching $10^{-2} \, M_\odot$ at 1,100 d.
In contrast, clumpy ejecta form $10^{-3} \, M_\odot$ of $\rm Mg_2SiO_4$ at 100 d and reach $10^{-2} \, M_\odot$ by 300 d with carbonaceous dust not forming until 700 d.
In all cases, $\rm Mg_2SiO_4$ dominates the total dust mass for the first 200-300 d of dust formation.  
In \cite{sluder2018}, another MNT model specifically for SN\,1987A, which can be compared to the clumpy 19 $M_\odot$ model of \cite{sarangi2015}, predicts 0.1 $M_\odot$ of $\rm Mg_2SiO_4$ by 300 d, with $10^{-2}\, M_\odot$ of carbonaceous dust by 500 d.
We present this range of models to demonstrate that the predictions from different dust formation models still widely disagree.
Broadly, the dust mass inferred from the SED of SN\,2017eaw of $10^{-4} (10^{-6}) \, M_{\odot}$ assuming silicate (carbonaceous) dust, starting at 200 d is a factor of 100 to 10,000 smaller than model predictions for a normal SN II-P. 


The small amount of dust determined from the IR SED fitting at early time is in line with dust mass determined from IR observations in other II-P SNe (e.g. \citealp{szalai2011,tinyanont2016}), including SNe\,2004dj and 2004et \citep{meikle2011, kotak2009}.
While the total amount of dust found at $\sim$few hundred days post explosion is insufficient for SNe II-P to be a major source of cosmic dust production, observations of SN\,1987A and Galactic SN remnants reveal 0.1-1 $M_\odot$ of dust.  
As shown by ALMA observations, SN\,1987A has as much as 0.2 $M_{\odot}$ of dust in its inner ejecta in 2012, 26 years after the explosion \citep{indebetouw2014}.
Long wavelength observations of supernova remnants also reveal similar amount of dust surviving the passage of the reverse shock, available to be dispersed into the ISM.
One explanation for this behavior is that dust continues to form in the ejecta over the course of few years, as predicted by \cite{sarangi2015}.
Another possibility is that dust is formed quickly, but is optically thick at early time.
As a result, the dust mass inferred from near to mid-IR observations at these epochs only accounts for dust in the outermost layer of the ejecta \citep{dwek2018}.
Future observations, especially at very late times, of nearby CCSNe are still needed to probe the evolution of dust from few hundreds to few thousands days post explosion, which will test different chemical evolution models of SN ejecta. 
Such observations will be greatly enabled in the era of \textit{James Webb Space Telescope}. 

\section{Summary and conclusions}\label{sec:discussion}
In this paper, we have presented the following:
\begin{itemize}
\item We presented observations of SN\,2017eaw's progenitor in the Ks band from 344 to $\sim$1 day before the explosion. 
We detected no photometric variability of the progenitor in this band down to $\Delta \nu L_\nu \lesssim 5{,}000 \, L_{\odot}$, 6\% of the $Ks$ band luminosity. 
Further, there was no evidence of short term variability $\sim$days before core-collapse.

\item SN\,2017eaw is similar in near-IR photometric evolution to normal SNe II-P 2002hh and 2004et, both also in NGC\,6946, although SN\,2017eaw is dimmer than SN\,2004et by 0.5 mag in the \textit{Spitzer}/IRAC 3.6 and 4.5 $\rm \mu$m bands. 
The \textit{Spitzer}/IRAC 3.6 and 4.5 $\rm \mu$m bands light curves of SN\,2017eaw and SN\,2004et are nearly identical from the time of the explosion out to our last epoch at 566 d. 
Long term monitoring of SN\,2017eaw in the mid-IR will reveal whether it will have a late-time CSM interaction that rebrightens the mid-IR light curve as was observed in SN\,2004et.

\item The SED evolution showed a rising warm dust emission continuum in the 3.6 $\mu$m band and an excess in the 4.5 $\mu$m band that was likely due to CO fundamental band emission. 
Small amounts of dust ($10^{-6} \, M_{\odot}$ carbonaceous or $10^{-4} \, M_{\odot}$ silicate) were needed to fit the SED at 197 and 295 d. 
While we cannot distinguish the two dust compositions from our data due to the lack of wavelength coverage around the silicate features at $\sim$10 $\mu$m, we note that silicate dust was detected in SN\,2004et. 

\item The CO first overtone band at 2.3 $\mu$m continued to evolve from 205 d as reported by \cite{rho2018}. 
The CO feature peaked at 389 d and subsequently faded with respect to the continuum in later epochs, similar to the behavior observed in SN\,1987A.  
We conclude that the CO feature faded because the gas had cooled enough so that CO vibrational levels required for the first overtone emission were no longer populated. 
The 4.5 $\mu$m excess, likely due to the CO fundamental bands, was still detected at 566 d.
The formation and survival of CO are in line with predictions from the MNT chemical evolution models presented by \cite{sarangi2013, sarangi2015, sluder2018}.

\item  At 389 d, the CO line profile was similar to that of SN\,1987A at 377 d.
As a result, a non-LTE model fitted to SN\,1987A \citep{liu1992} also fits our data. 
This indicates that the CO mass ($\sim 10^{-3}\, M_{\odot}$) and temperature (1,800 K) of the two SNe are similar at this epoch. 



\end{itemize}

This study underlines the need for future IR observations of CCSNe in order to study molecule and dust formation in their ejecta in greater detail. 
Such observations, enabled by advances in IR instruments, especially \textit{JWST}, will allow us to put constraints on chemical models of SNe ejecta and to paint a more complete picture of CCSNe's contribution to the molecular and dust budget of the interstellar medium.

\acknowledgments 
We thank Jim Fuller for helpful discussions and input on the paper draft. 
Some of the data presented herein were obtained at the W. M. Keck Observatory, which is operated as a scientific partnership among the California Institute of Technology, the University of California and the National Aeronautics and Space Administration. The Observatory was made possible by the generous financial support of the W. M. Keck Foundation.
The authors wish to recognize and acknowledge the very significant cultural role and reverence that the summit of Maunakea has always had within the indigenous Hawaiian community.  
We are most fortunate to have the opportunity to conduct observations from this mountain.
Some of the data presented herein were obtained at Palomar Observatory, which is operated by a collaboration between California Institute of Technology, Jet Propulsion Laboratory, Yale University, and National Astronomical Observatories of China. 
This work is based in part on observations made with the \textit{Spitzer} Space Telescope, which is operated by the Jet Propulsion Laboratory, California Institute of Technology under a contract with NASA. Support for this work was provided by NASA through an award issued by JPL/Caltech.
This research has made use of the NASA/IPAC Extragalactic Database (NED), which is operated by the Jet Propulsion Laboratory, California Institute of Technology, under contract with the National Aeronautics and Space Administration.
This research made use of Astropy, a community-developed core Python package for Astronomy \citep{astropy2018}.
RDG was supported by NASA and the United States Air Force.

\facilities{Hale (WIRC, TripleSpec), Spitzer, Keck (MOSFIRE, NIRES)}

\bibliography{SN2017eaw.bib}

\end{document}